\documentclass[iop]{emulateapj}
\slugcomment{Accepted to ApJ: 23 Oct. 2014}

\usepackage[backref,breaklinks,colorlinks,citecolor=blue]{hyperref}
\usepackage[all]{hypcap}

\usepackage{natbib}
\usepackage{graphicx} 
\usepackage{mathrsfs} 
\usepackage{amsmath}
\usepackage{float}
\usepackage{amssymb}
\usepackage{rotating}

\newcommand{\galex}{{\it GALEX}}
\newcommand{\ha}{H$\alpha$}

\usepackage[symbol*]{footmisc}
\DefineFNsymbolsTM{myfnsymbols}{%
  \textdagger \dagger
  \textdaggerdbl   \ddagger}%
\setfnsymbol{myfnsymbols}

\begin{document}

\title{The Near-Ultraviolet Luminosity Function of Young, Early M-Type Dwarf Stars}

\author{Megan Ansdell\altaffilmark{1}, Eric Gaidos\altaffilmark{2},
  Andrew W. Mann\altaffilmark{3}\footnote{Harlan J. Smith Postdoctoral Fellow}, Sebastien
  L\'{e}pine\altaffilmark{4,5}, David James\altaffilmark{6}, Andrea
  Buccino\altaffilmark{7,8}, Christoph Baranec\altaffilmark{1},
  Nicholas M. Law\altaffilmark{9}, Reed Riddle\altaffilmark{10},
  Pablo Mauas\altaffilmark{7}, Romina
  Petrucci\altaffilmark{7}\footnote{Visiting Astronomer, Complejo
    Astron\'{o}mico El Leoncito, operated under agreement between the
    Consejo Nacional de Investigaciones Cient\'{i}ficas y T\'{e}cnicas de la
    Rep\'{u}blica Argentina and the National Universities of La Plata,
    C\'{o}rdoba, and San Jaun}}
\altaffiltext{1}{Institute for Astronomy, University of Hawaii, Honolulu, HI 96822}
\altaffiltext{2}{Department of Geology \& Geophysics, University of Hawaii, Honolulu, HI 96822}
\altaffiltext{3}{Department of Astronomy, University of Texas at Austin, Austin, TX 78712}
\altaffiltext{4}{Department of Physics \& Astronomy, Georgia State University, Atlanta, GA 30302}
\altaffiltext{5}{Department of Astrophysics, American Museum of Natural History, New York, NY 10024}
\altaffiltext{6}{Cerro Tololo Inter-American Observatory, Casilla 603, La Serena, Chile}
\altaffiltext{7}{Instituto de Astronom\'{i}a y F\'{i}sica del Espacio, C1428EHA - Buenos Aires, Argentina}
\altaffiltext{8}{Departamento de F\'{i}sica, FCEN-Universidad de Buenos Aires, Argentina}
\altaffiltext{9}{Department of Physics and Astronomy, University of North Carolina at Chapel Hill, Chapel Hill, NC 27599}
\altaffiltext{10}{Division of Physics, Mathematics, and Astronomy, California Institute of Technology, Pasadena, CA, 91125}

\begin{abstract}

Planets orbiting within the close-in habitable zones of M dwarf stars
will be exposed to elevated high-energy radiation driven by strong
magneto-hydrodynamic dynamos during stellar youth. Near-ultraviolet
(NUV) irradiation can erode and alter the chemistry of planetary
atmospheres, and a quantitative description of the evolution of NUV
emission from M dwarfs is needed when modeling these effects.
We investigated the NUV luminosity evolution of early M-type
dwarfs by cross-correlating the \citet{2011AJ....142..138L} catalog of
bright M dwarfs with the \galex{} catalog of NUV (1771--2831\AA) sources. Of the
4805 sources with \galex{} counterparts, 797 have NUV emission
significantly ($>2.5\sigma$) in excess of an empirical basal
level. We inspected these candidate active stars using visible-wavelength
spectra, high-resolution adaptive optics imaging, time-series
photometry, and literature searches to identify cases where the
elevated NUV emission is due to unresolved
background sources or stellar companions; we estimated the overall
occurrence of these ``false positives'' as $\sim$16\%. We constructed
a NUV luminosity function that accounted for false positives,
detection biases of the source catalogs, and \galex{} upper limits. We
found the NUV luminosity function to be
inconsistent with predictions from a constant star-formation rate and
simplified age-activity relation defined by a two-parameter power law.

\end{abstract}

\section{INTRODUCTION} \label{sec-intro}

The characteristics of M dwarf stars make them favorable targets
in the search for Earth-like planets. Their ``habitable
zones'' (i.e., the range of orbital semi-major axes at which liquid
water is stable on an Earth-like planet) are more compact than those
of solar-type stars due to their comparatively low
luminosities. These closer orbits make Earth-like planets possible to
detect with radial velocity and transit methods
\citep{2007AsBio...7...30T,2007AsBio...7...85S,2007Sci...318..210G}. 

However, planets in the close-in habitable zones of M dwarfs may
be exposed to elevated levels of high-energy radiation. The photospheres of M dwarfs
emit negligibly at short wavelengths due to low
effective temperatures and absorption by neutral iron (Fe I). Yet
these stars can emit ultraviolet (UV) and X-ray radiation from
their upper chromospheres, coronae, and active regions due to heating from their
strong magnetic-hydrodynamic dynamos. 

UV radiation may play opposing roles in planet habitability
and the origins of life. Elevated UV radiation can irreparably damage
organisms on planetary surfaces as well as erode planetary atmospheres by
injecting heat that drives escape of hydrogen and other volatiles required for life
\citep[e.g.,][]{2009ApJ...703..905T,2013AsBio..13.1011E,2014arXiv1410.2112M}.  On the
other hand, atmospheric UV photochemistry is a potential source of
prebiotic molecules \citep{2002ESASP.518....9E} and UV radiation can
induce mutations upon which natural selection can act
\citep[e.g.,][]{JEU:JEU548}. Buccino et al. 2006 defined the
``UV habitable zone'' as the distance at which a planet is close
enough to receive sufficient UV radiation to enable biogenesis processes, but also
far enough to avoid irreparable damage to DNA by exposure to
heightened levels of UV flux. Buccino et al. 2007 derived the UV habitable zones
for three planet-hosting M dwarfs (GJ 581, GJ 849, GJ 876) with UV spectra
from the {\it International Ultraviolet Explorer} (IUE). They found that for all three systems
the liquid-water and UV habitable zones did {\it not}
overlap, and suggested that an alternative source of UV emission, such as
stellar flares, might be needed to enable prebiotic chemistry
on planets orbiting in the liquid-water habitable zones around M dwarfs.

The UV emission of M dwarfs is known to evolve
with age. Young M dwarfs ($\lesssim100$ Myr) exhibit rapid rotation
with strong magneto-hydrodynamic dynamos that result in enhanced UV (as well as X-ray) emission.
As stars age, angular momentum is gradually lost through stellar winds,
causing stars to spin down and become less active with time.  Single
early-M dwarfs (M0-M3) remain rapidly rotating and active for $\sim$1
Gyr, while many late-M dwarfs (M5-M9) stay in this state for up to
$\sim$8 Gyr \citep{2008AJ....135..785W,2013MNRAS.429.3570R}. 
This dichotomous behavior may result from the appearance of
fully convective interiors and low Rossby numbers near the M4 spectral
subtype (\citealt{lrsp-2012-1,2013A&A...549L...5G}; see also
\citealt{2012ApJ...746...43R} for an alternative explanation for this
dichotomy). Due to survey biases, the vast majority of known
planet-hosting M dwarfs have early spectral subtypes. A quantitative
description of the evolution of UV emission from early-M dwarfs is
therefore important for accurately modeling the effects of UV irradiation on
planets orbiting these stars \cite[e.g.,][]{2014arXiv1410.2112M}. 
Moreover, future exoplanet surveys---such as K2, the Transiting
Exoplanet Survey Satellite \citep[TESS;][]{2014ebi..conf.3.10R}, and the Next Generation
Transit Survey \cite[NGTS;][]{2013EPJWC..4713002W}---will continue to monitor early-M dwarfs. These
surveys will generate substantial data on the variability
and rotation of these stars, which can then be compared to their UV emission.

\begin{figure}
\begin{centering}
\includegraphics[width=8.5cm]{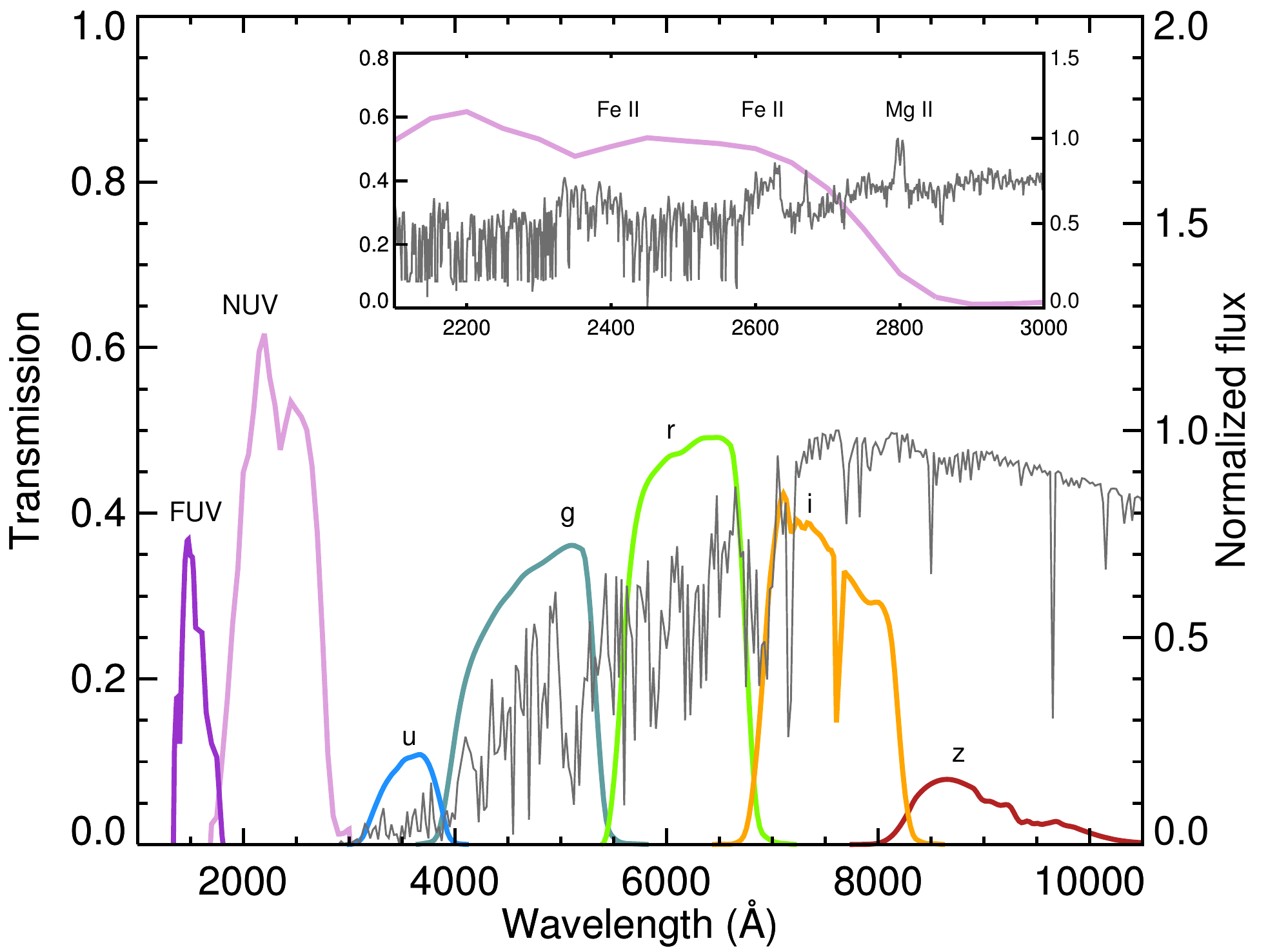}
\caption{\small \galex{} FUV and NUV bandpasses compared to SDSS 
  optical filters ($ugriz$). A PHOENIX model spectrum of an M dwarf
  with $T_{eff}=3700$ K and solar metallicity is plotted for
  reference (gray curve). The inset shows the NUV spectrum of an active,
  planet-hosting M dwarf \citep[GJ 876;][]{2013ApJ...763..149F} with
  the GALEX NUV bandpass (not to same scale) and activity-related
  emission lines shown for reference.}
\label{fig-filt}
\end{centering}
\end{figure}

Much of the UV emission from astronomical objects must be observed from space due
to absorption by the Earth's atmosphere at these short
wavelengths. The Galaxy Evolution Explorer (\galex{};
\citealt{2005ApJ...619L...1M}) is a recently decommissioned
space-based telescope that performed all-sky
imaging in both near-UV (NUV; 1771--2831\AA) and far-UV (FUV;
1344--1768\AA) bandpasses (Figure~\ref{fig-filt}). Several recent
studies have used \galex{} data to identify active and/or young M dwarfs
\citep{2009PASP..121..450B,2011ApJ...727...62R,2011ApJ...727....6S,2013ApJ...774..101R,2013MNRAS.431.2063S}.
Most notably, \citet{2013MNRAS.431.2063S} studied a volume-limited
sample of 159 field M dwarfs within 10 pc, which they identified by cross-correlating
the \citealt{2011AJ....142..138L} (hereafter LG11) catalog of
bright M dwarfs with the \galex{} sixth data release (GR6). They
compared their sample to members of the TW Hydra young moving group
(YMG), which has a known age of $\sim$10 Myr, to derive a power-law
age-activity relation. They found that the UV luminosities of early-M
dwarfs decline by roughly three orders of magnitude from $\sim$10 Myr
to a few Gyr of age. 

However, the \citet{2013MNRAS.431.2063S} sample was too small to
quantitatively describe a UV luminosity function, particularly at
higher UV luminosities where there are fewer stars. Furthermore, M
dwarfs can appear UV luminous for reasons other than stellar
youth. Such reasons include: unrelated
background UV sources confused in the $\sim$5 arcsec beam of \galex{};
companion white dwarfs or late-M dwarfs with persistent UV emission;
or tidally locked binaries in which spin-orbit
synchronization induces ongoing activity. These ``false
positives'' (FPs) must be identified and removed, at
least in a statistical sense, in order to estimate a UV
luminosity function.

In this work we utilize the entire LG11 catalog of bright M dwarfs,
cross-correlating it with the final version of the \galex{} all-sky UV source catalog
(GR7), to derive a NUV luminosity function (NUVLF) for early-M dwarfs (M0-M3). We
describe our sample selection in Section \ref{sec-sample} and detail our
follow-up observations in Section \ref{sec-data}. In Section
\ref{sec-fp} we describe our methods for identifying FPs and then
estimate the overall FP rate in our sample using a maximum likelihood
method. We construct the NUVLF in Section \ref{sec-analysis} by
accounting for FPs, the detection biases of the source
catalogs, and \galex{}  upper limits. In Section \ref{sec-diss}, we
compare the NUVLF to simple models of star formation and M dwarf
activity evolution as well as describe the implications
and caveats of our findings.

\capstartfalse
\capstartfalse

\section{SAMPLE} \label{sec-sample}

\subsection{Identifying M Dwarfs in the \galex{} Catalog} \label{sec-uvsample}

We cross-correlated the LG11 catalog of 8889 nearby (${\rm d}\lesssim60$ pc),
bright ($J<10$), K7--M5 stars with the final
\galex{} data release (GR7). We searched the \galex{} All-sky Imaging
Survey (AIS), which contains both NUV and FUV
sources, but kept only NUV matches as \galex{} was much more sensitive
in this bandpass (e.g., a preliminary cross-correlation by LG11
found five times more matches to NUV than FUV sources).

Our cross-correlation included a correction for stellar proper motion
between the J2000 epoch of LG11 and the epochs of the individual \galex{}
AIS observations. We used the {\it GalexView} online
tool\footnote{http://galex.stsci.edu/GalexView/} to perform a preliminary cross-correlation
between the LG11 and \galex{} AIS catalogs with a 1 arcmin match
radius. This search radius accounted for the resolution of \galex{} ($\sim$5
arcsec) plus the maximum proper motion of any LG11 star over 10 years
(roughly the time between the J2000 epoch and the last \galex{} AIS
observation). This preliminary search returned multiple \galex{}  AIS
source matches for most LG11 stars, as expected from the large search
radius. For each LG11 star we calculated its expected position on the sky
at the observation date of each of its \galex{}  AIS matches. We then re-performed the
cross-correlation using the adjusted LG11 positions and a reduced
matching criterion of 5 arcsec (the resolution of \galex{}) to identify the final
source matches.

\begin{figure}
\begin{centering}
\includegraphics[width=8.5cm]{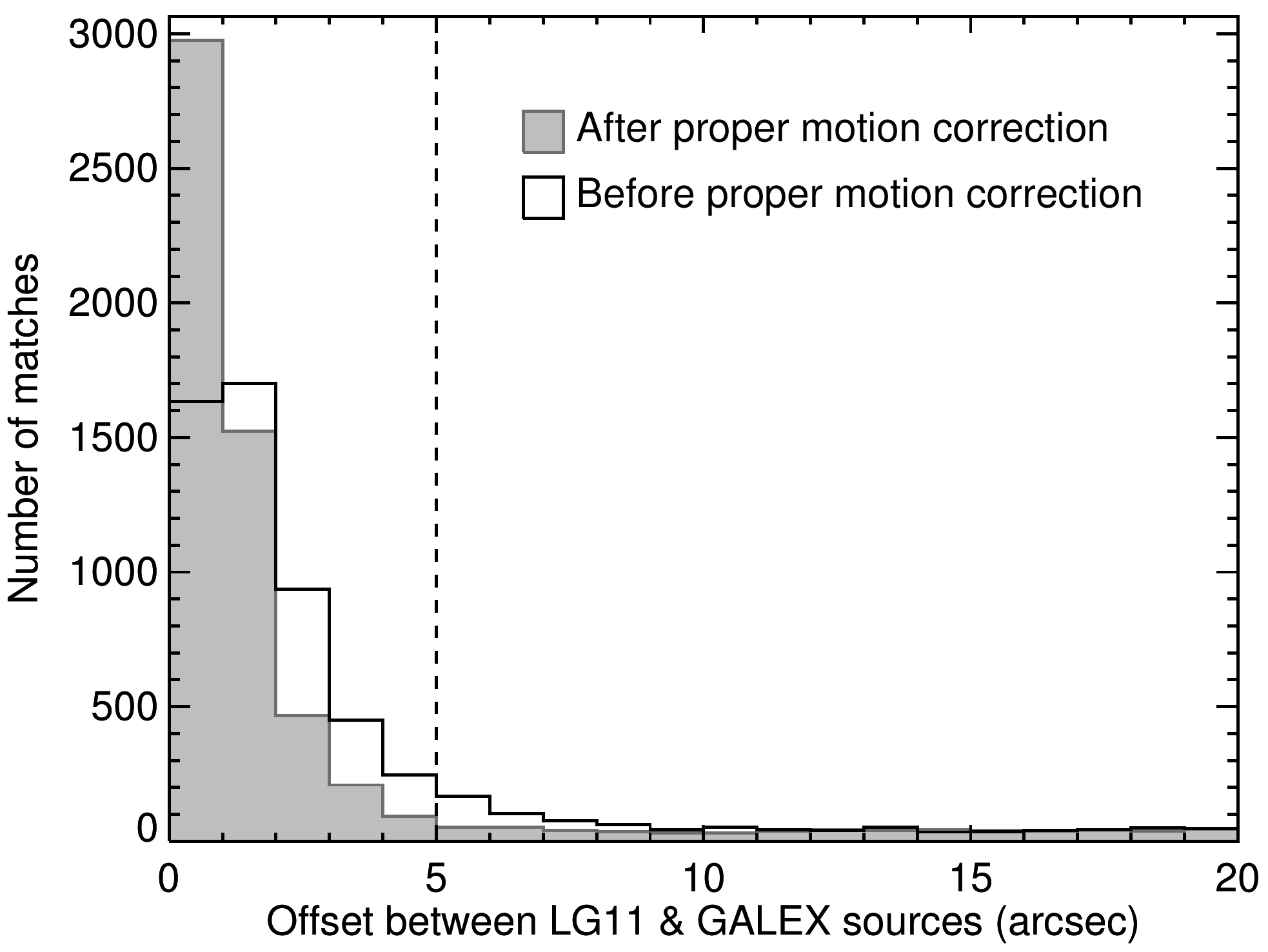}
\caption{\small Improved cross-correlation between LG11 and {\it
    GALEX} AIS catalogs after proper motion correction. The dashed line
  shows our match criteria of $\lesssim$ 5 arcsec separation
  (corresponding to the angular resolution of \galex{}). After proper
  motion correction, the number of matches increased by $\sim$300 and the 
  residual angular separations significantly decreased, as shown by the shift toward smaller offsets.}
\label{fig-offset}
\end{centering}
\end{figure}

There were 1251 LG11 stars with multiple matches to the
\galex{} AIS catalog, even after correcting for proper motion and
applying the stricter 5 arcsec search radius. For these we simply took the closest
match. This was justified because almost all of these multiple matches were due to
repeated \galex{} AIS observations of a given area of sky: the matches had
different \galex{} AIS tile numbers and/or different exposure times,
and thus are likely the same star. However there were 49 multiple matches
that had the same \galex{} tile number and exposure time, therefore
representing the case when two objects are close enough on the sky
to be confused by \galex{}. The NUV magnitude
differences for these 49 multiple matches were only $\sim$0.30 on average, and
just 8 of these sources were ultimately used in the NUVLF. Thus the
choice of which source to match should not significantly affect our
derived NUVLF. Figure~\ref{fig-offset} compares the cross-correlation results before
and after proper motion correction: the proper motion correction
significantly improved the cross-correlation results, adding
$\sim$300 new matches compared to the number of matches prior to
proper motion correction, and significantly decreasing the residual angular
separations of the matches, as illustrated by the histogram shift
toward smaller offsets. 

This cross-correlation process identified 5267 LG11 stars with NUV counterparts
in the \galex{} AIS catalog. Figure~\ref{fig-uv} plots $m_{\rm
  NUV}-K_{\rm S}$, a distant-independent measure of NUV luminosity,
vs. $V-J$, a proxy for stellar effective temperature, for this sample. $m_{\rm NUV}$ is from the
\galex{} AIS catalog, $J$ and $K_{\rm S}$ are from the Two Micron All Sky Survey
\citep[2MASS;][]{2006AJ....131.1163S}, and $V$ is from the AAVSO
Photometric All-Sky Survey \citep[APASS;][]{2012JAVSO..40..430H}, the
Tycho-2 and {\it Hipparcos} catalogs \citep{1997ESASP1200.....P}, or
generated from USNO-B magnitudes \citep{2005AJ....129.1483L}.

\capstartfalse
\capstartfalse

\subsection{Determining the Basal NUV Locus} \label{sec-basal}

Most sources in Figure~\ref{fig-uv} fall along a locus
(designated by gray points according to the criterion described below), which we
interpret as the basal level of NUV emission for early-M dwarfs. To
describe this basal NUV emission as a function of
stellar effective temperature, we fit a line to median values of
$m_{\rm NUV}-K_{\rm S}$ vs. $V-J$ using
only stars with errors \textless10\% in all bands and $V-J$ colors
$<3.5$. We iteratively removed outlier stars and re-performed the
fit until the remaining stars were Gaussian-distributed in $m_{\rm
  NUV}-K_S$ about the median-fit line. This gave a best-fit
(designated by the black dashed line in Figure~\ref{fig-uv}) with the following
parameters:

\begin{equation}
m_{\rm NUV}-K_{\rm S} = 7.72 + 1.66 (V-J)
\label{eqn-locus}
\end{equation}

This basal NUV locus is presumably an extension of the locus of inactive
solar-type stars, which reaches $m_{\rm NUV}-K_{\rm S}\sim 10$ at $V-J
\sim 2$ \citep[see Figure 2 in][]{2011AJ....142...23F}.  However, our
basal locus is much bluer than predicted by PHOENIX
model spectra \citep{2013MSAIS..24..128A,2013A&A...556A..15R} using
solar abundances \citep{2011SoPh..268..255C}
and log $g  = 5.0$. Thus this locus likely does {\it not} exclusively represent
photospheric NUV emission. It also cannot be an artifact of a constant \galex{}
flux limit, as the lower right-hand region of Figure \ref{fig-uv}
is populated while the lower left-hand region is not. Instead, this locus
likely indicates persistent NUV line emission from a
higher-temperature upper chromosphere. \citet{2013MNRAS.431.2063S}
previously noted that all M dwarfs appear to exhibit NUV emission in excess of their
expected photospheric value from stellar atmosphere models.

\subsection{Selecting the NUV-Luminous Stars} \label{sec-bright}

\begin{figure}
\begin{centering}
\includegraphics[width=8.5cm]{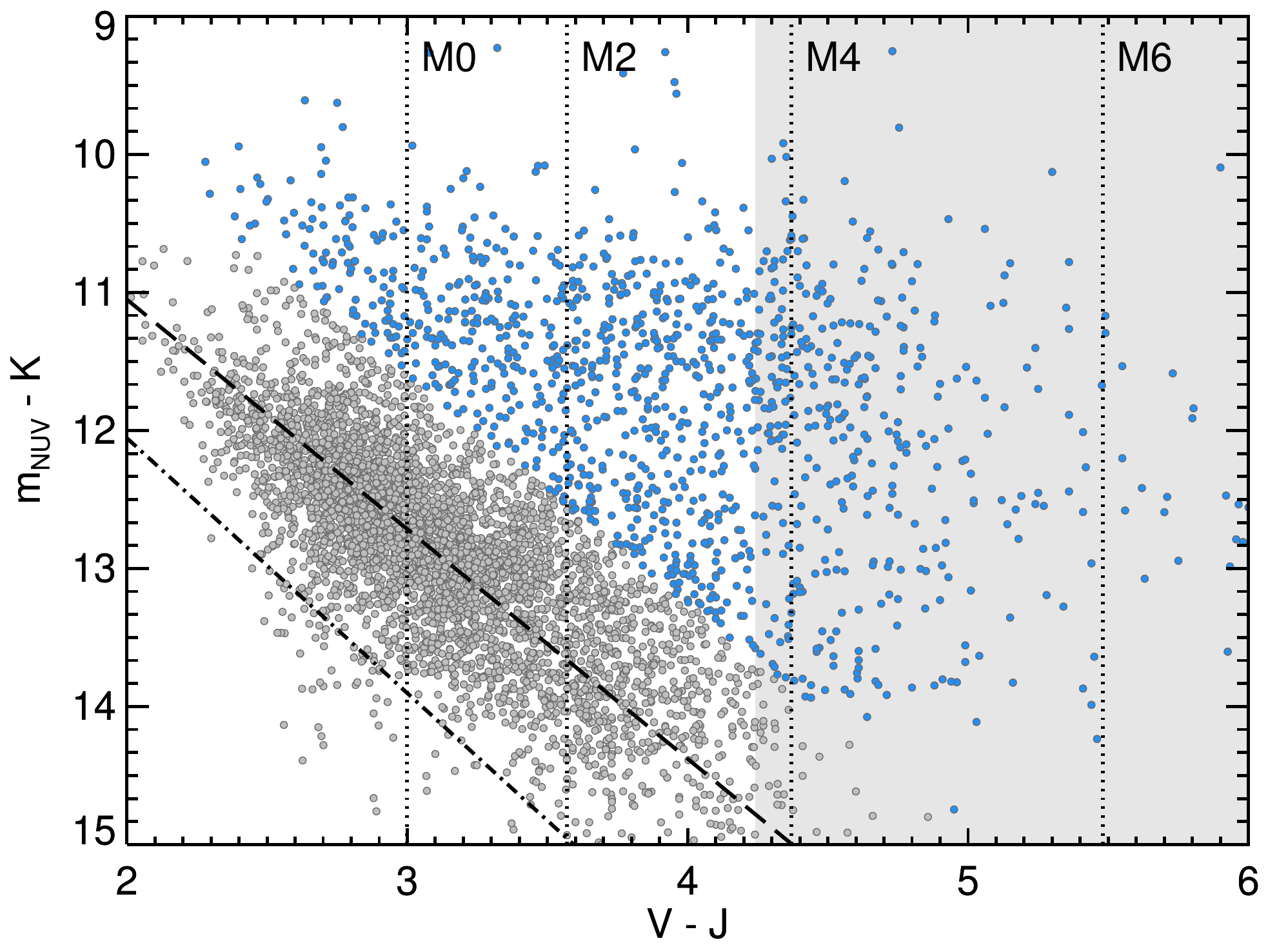}
\caption{\small LG11 stars with NUV counterparts in the \galex{} AIS
  catalog. The $y$-axis is a measure of NUV luminosity and the
  $x$-axis is a proxy for stellar effective temperature that
  correlates to spectral subtype. The dominant locus (gray points)
  likely represents older M dwarfs, while the distinct population
  toward higher NUV luminosities (blue points) presumably
  represents young/active stars. Approximate spectral subtypes
  are shown for reference; stars later than M3 (indicated by the
  shaded region) were ultimately removed from our sample (see Section
  \ref{sec-latem}). The black dashed-dotted line shows the predicted
  photospheric level of NUV emission from PHOENIX model spectra. The
  dominant locus of older stars exhibits higher NUV emission than the predicted
  photospheric values, suggesting that all M dwarfs exhibit a basal
  level of NUV emission from their chromospheres. The median fit
  to this basal emission level is shown by the black dashed line and
  given in Equation~\ref{eqn-locus}.}
\label{fig-uv}
\end{centering}
\end{figure}

The distribution in Figure~\ref{fig-uv} also features a smaller
population of stars with NUV emission significantly in excess of the
empirically determined basal value. We interpret these NUV-luminous sources as being
mostly young stars exhibiting heightened activity, but also including ``false
positives'' that appear active for reasons other than stellar
youth. We identified the 1210 NUV-luminous stars from the 5267 NUV-detected stars as those with
$m_{\rm NUV}-K_{\rm S}$ colors at least $2.5\sigma$ ($\sim$1.12
magnitudes) bluer than their expected basal value given by
Equation~\ref{eqn-locus}. We identify these stars as blue points in Figure~\ref{fig-uv}.

\capstartfalse
\capstartfalse

\subsection{Removing Late-M Dwarfs} \label{sec-latem}

We only consider early-M dwarfs (M0-M3) in the remainder of this
work. This is because we identify our sample of young M dwarfs by their heightened
UV luminosity (see Section~\ref{sec-bright}), a property that has been
observed to evolve significantly with stellar age for early-M dwarfs
and at a much slower rate for late-M dwarfs (M5-M9). Namely, observations
indicate that early-M dwarfs are rapidly
rotating and active for only $\sim$1 Gyr, while late-M dwarfs
remain in this state for up to $\sim$8~Gyr
\citep{2008AJ....135..785W,2013MNRAS.429.3570R}. This difference may
be due to the transition to fully convective interiors around the M4
spectral subtype \citep{lrsp-2012-1,2013A&A...549L...5G}. We therefore
removed stars with spectral subtypes later than M3 as a first
step to ensuring our measured NUV excess is due
to stellar youth rather than a longer duration of stellar activity.

To identify late-M dwarfs we used the empirical relation between
$V-J$ color and spectral subtype derived in LG11 and then revised in \cite{2013AJ....145..102L}
(see their Equation 12). These color-based assignments provide only rough estimates ($\pm$
1 spectral subtype), however they also provide a uniform method of spectral
classification, which is important for statistical studies such as
this. Moreover, a rough estimate is sufficient because we are using these
spectral subtypes to identify late-M dwarfs and the spectral
subtype boundary between early- and late-M dwarfs is not well defined (i.e., somewhere
between M3 and M5). 

Removing the late-M dwarfs from our original sample of NUV-detected M
dwarfs (see Section \ref{sec-uvsample}) resulted in a final sample of 4805 NUV-detected
early-M dwarfs of which 797 were identified as NUV-luminous (see
Section \ref{sec-bright}). All references hereafter to our ``sample''
are to this final selection that includes only
early-M dwarfs, unless explicitly stated otherwise. Parameters of
the NUV-luminous sources are presented in Table~\ref{tab-params}.

\subsection{Characterizing the Basal NUV Locus} \label{sec-quiescent}

Figure~\ref{fig-uv} suggests that the locus of stars exhibiting
basal NUV emission (gray points; see Section~\ref{sec-uvsample}) has
an intrinsic width. We estimated this intrinsic width by first taking the standard deviation of
distances in $m_{\rm NUV}-K_{\rm S} $ from the locus median (black dashed line) to
all stars in the locus with errors \textless10\% in all bands. We accounted for
measurement errors by subtracting in quadrature the median measurement
errors for each band: the $m_{\rm NUV}$ errors were taken from the
\galex{} AIS catalog, the $V$-band errors were taken from the respective catalogs
(see Section \ref{sec-uvsample}), and the $K_{\rm S}$- and $J$-band errors
were considered negligible at \textless1\%. Errors in $V-J$ were translated
to $m_{\rm NUV}-K_{\rm S}$ using the slope of our median fit to the locus (see
Equation~\ref{eqn-locus}) before subtracting in
quadrature. We found that the locus of stars exhibiting basal NUV
emission in Figure~\ref{fig-uv} has an intrinsic width of $\sim$0.50
magnitudes in $m_{\rm NUV}-K_{\rm S}$,
after accounting for measurement errors.

The locus width could be the product of several factors, in particular
stellar variability, interstellar extinction, unresolved binaries,
and metallicity variations. We first investigated stellar
variability using the 1202 LG11 stars in our sample with multiple
matches to the \galex{} AIS catalog that were likely repeated
observations of the same star at different epochs rather than source
confusion (see Section \ref{sec-uvsample}). We found the maximum difference in
NUV magnitude for each multiple match, then took the standard
deviation as an estimate of the error introduced by stellar
variability. We then subtracted this value ($\sim$$0.24$) in quadrature along with
the measurement errors (see above), which
reduced the estimated intrinsic locus width by only $\sim$0.07
magnitudes. This shows that stellar variability is not a
significant contributor to the locus width. 

We also confirmed that the locus width was not due to interstellar extinction,
$A(V)$. UV sources are subject to significant interstellar extinction,
as evident by the significant drop in \galex{} detections near the
Galactic plane \citep[see Figure 1
  in][]{2011MNRAS.411.2770B}. However, interstellar extinction should
be negligible for our sample as most LG11 stars reside within 60
pc and are therefore contained within a low-density
($\sim$0.005 atoms cm$^{-3}$) region known as the ``Local Bubble"
\citep{1987ARA&A..25..303C}. Nevertheless, we tested whether
interstellar extinction could account for the
locus scatter by searching for a minimum locus width as a function of
assumed extinction per parsec, as described below. We found distances
to each star using the $J$-band photometric distance,
where M$_{J}$ was obtained from $V-J$ color (although LG11
computes photometric distances, we re-compute them here using an
updated color magnitude relation given by Equation 22 in
\citealt{2013AJ....145..102L}). We used extinction coefficients from
\cite{2013MNRAS.430.2188Y}, which produced reddening corrections of
E$_{(NUV-K)}=2.24\times A(V)$ and E$_{(V-J)}=0.77\times A(V)$. We
tested $A(V)$ values ranging from 0 to 0.001 mag pc$^{-1}$ at a
cadence of $1.5\times10^{-6}$ mag pc$^{-1}$. This encompassed values well beyond
the expected interstellar extinction within 60 pc; assuming $A(V)
\sim1$ mag kpc$^{-1}$ along the Galactic plane, and conservatively
assuming a Local Bubble that is 10\% the typical density of the
interstellar medium, the expected interstellar extinction within 60 pc
is $A(V)\sim0.0001$ mag pc$^{-1}$. We applied reddening corrections to
each star based on their individual distances, then re-measured the
locus width for each $A(V)$ test value. We found no local minimum in
the locus scatter. Rather, attempting to correct for extinction only
increased the scatter of the locus. 

We therefore concluded that stellar variability and interstellar
extinction do not contribute significantly to the locus
width. However, the locus width may be due to unresolved binaries,
metallicity-dependent stellar colors, and
continued variation of the basal NUV emission level with
age. Unfortunately our dataset did not allow us to investigate these
possibilities in detail.

\subsection{Comparing FUV \& X-ray Emission} \label{sec-other}

We checked for FUV and X-ray counterparts to our sample. To obtain FUV
counterparts we simply took the FUV sources associated
with our NUV matches to the \galex{} AIS catalog. To identify stars
with X-ray counterparts, we cross-correlated our sample with the {\it ROSAT}
All-Sky Survey Bright Source Catalog \citep{1999A&A...349..389V} and
Faint Source Catalog \citep{2000IAUC.7432....3V}.  We used a 25 arcsec
search radius around the LG11 coordinates, corresponding to the
2$\sigma$ {\it ROSAT} positional uncertainty determined by
\cite{1999A&A...349..389V}.  We converted the PSPC detector count rate
into an X-ray flux, F$_{\rm X}$, using the conversion factor from
\cite{1995ApJ...450..392S}: ${\rm CF}=(5.30{\rm HR}+8.31)10^{-12}$
ergs cm$^{-2}$ count$^{-1}$ where HR is the first hardness ratio from
the {\it ROSAT} catalog.  We did not correct for proper motion for
this cross-correlation due to the large positional uncertainty of
{\it ROSAT} compared to \galex{}. However we checked for
mismatches with background galaxies and quasars
by plotting F$_{\rm X}$/F$_{K_{\rm S}}$ as a function of $J-K_{\rm
  S}$, but found no significant outliers, i.e sources with F$_{\rm X}$/F$_{K_{\rm S}} \ge
  0.1$ \citep{2010MNRAS.405.2062K}. 

\begin{figure}
\begin{centering}
\includegraphics[width=8.5cm]{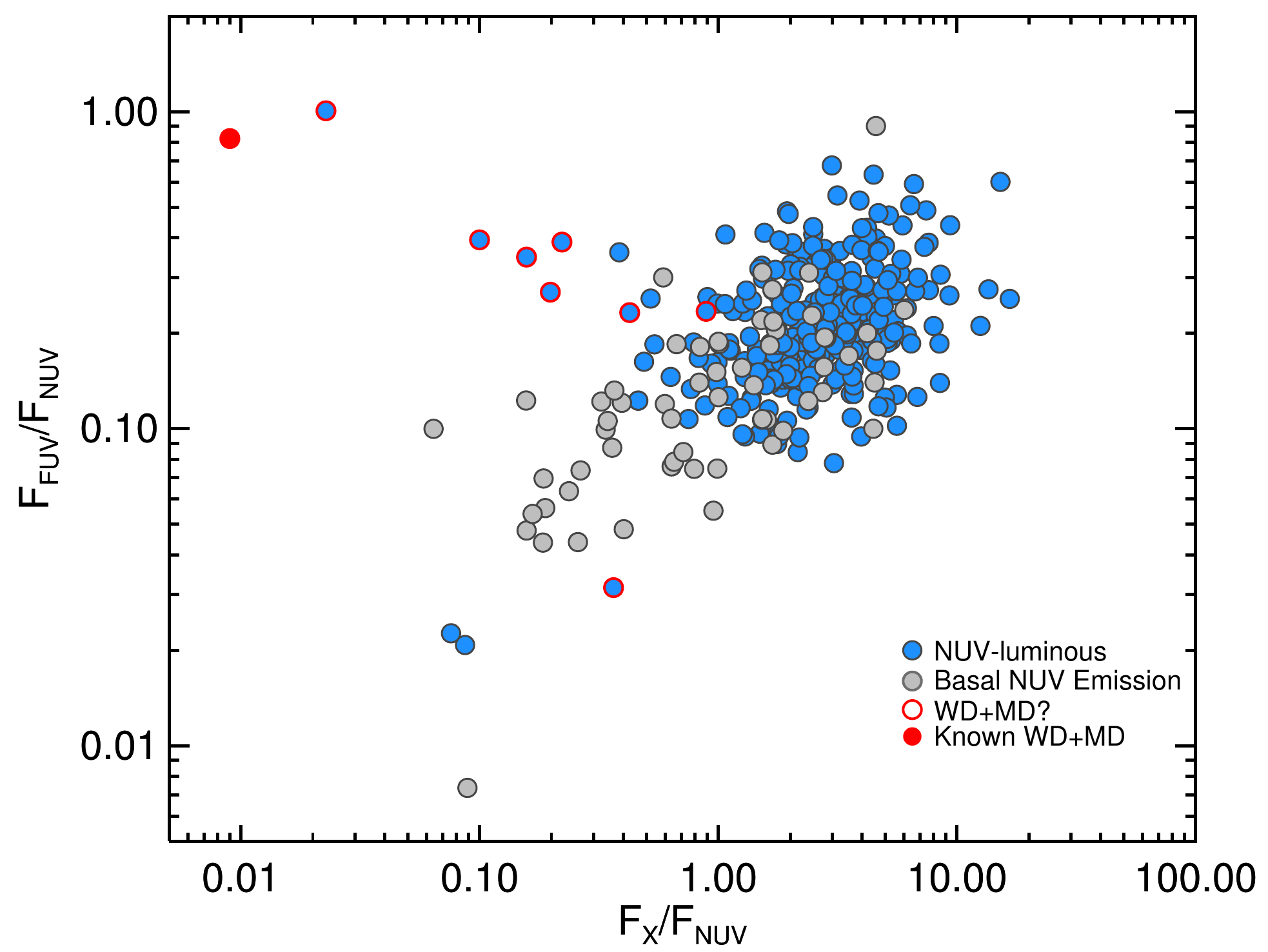}
\caption{\small The 387 NUV-detected early-M dwarfs in our sample that
  were also detected in {\it GALEX} FUV
  and {\it ROSAT} X-ray bands. Blue points correspond to NUV-luminous
  stars while gray points represent stars with basal NUV emission, as in Figure~\ref{fig-uv}.
  Candidate and known WD+MD systems are shown by open and filled red
  circles, respectively.}
\label{fig-other}
\end{centering}
\end{figure}

Only $\sim$$8\%$ of our sample (387 of 4805 sources) had
detectable flux in all three wavelength bands (NUV, FUV, and
X-ray). However $\sim$$85\%$ of this multi-wavelength subsample
(328 of 387 sources) was also selected as NUV-luminous in
Section~\ref{sec-bright}, which means $\sim$$40\%$ of our
NUV-luminous subsample (328 of 797 sources) was detected in all
three bands. Figure~\ref{fig-other} shows F$_{\rm FUV}/$F$_{\rm NUV}$ vs.
F$_{\rm X}/$F$_{\rm NUV}$ for the 387 multi-wavelength sources. Sources with ``hard''
spectra are located in the upper right while those with ``soft'' spectra
are located in the lower left. Sources with high FUV but low X-ray
emission could be M dwarfs with white dwarf companions (MD+WD pairs), as
white dwarfs emit strongly in the FUV due to their hot photospheres
but lack coronae from which X-ray emission typically
originates. Figure~\ref{fig-other} highlights a known WD+MD
pair and several MD+WD candidates identified by their high F$_{\rm NUV}/$F$_{J}$
ratios \citep[e.g., see Figure 3 in][]{2011ApJ...727....6S}.

\capstartfalse
\capstartfalse

\section{FOLLOW-UP OBSERVATIONS} \label{sec-data}

\subsection{Medium-resolution Optical Spectra} \label{sec-spectra}

We obtained medium-resolution ($\lambda/\Delta \lambda \sim 1000$)
optical spectra for 2128 out of the 4805 M dwarfs in our sample. The majority of
these (1307 spectra) were acquired using the Super-Nova Integral Field
Spectrograph, SNIFS (\citealt{2002SPIE.4836...61A}; \citealt{2004SPIE.5249..146L}),
mounted on the University of Hawaii 2.2-m telescope atop
Maunakea. SNIFS uses a dichroic mirror to separate incoming light into
blue (3200--5200\AA) and red (5100--9700\AA) spectrograph
channels. We only used spectra from the red channel as M dwarfs have very low
signal in the blue channel.  All spectra had signal-to-noise ratios
(${\rm SNR}$) $\gtrsim80$ per
resolution element in the red channel while avoiding the non-linear
regime of the detector. Details of our SNIFS data reduction method can
be found in \cite{2012ApJ...753...90M} and \cite{2013AJ....145..102L}. SNIFS is an
integral field spectrograph and therefore also provides limited
spatial information in the form of image cubes. A SNIFS image cube
covers 6 arcsec $\times$ 6 arcsec at 0.4 arcsec per pixel.

The remaining 821 spectra were obtained using four instruments on
three different telescopes: the Mark III spectrograph and the Boller
\& Chivens CCD spectrograph (CCDS) on the 1.3-m McGraw-Hill telescope
at the MDM Observatory on Kitt Peak (564 spectra); the RC spectrograph
on the 1.9-m Radcliffe telescope at the South African Astronomical
Observatory (SAAO) in South Africa (67 spectra); and the REOSC
spectrograph on the 2.15-m Jorge Sahade telescope at the Complejo
Astron\'{o}mico El Leoncito Observatory (CASLEO) in Argentina (190
spectra). Details of the data reduction methods for these spectra are
in \cite{2014MNRAS.443.2561G}.

\subsection{Robo-AO High-resolution Imaging} \label{sec-robo}

We observed 193 M dwarfs in our sample with the Robo-AO laser adaptive
optics and imaging system (\citealt{2013JVE....7250021B};
\citealt{2041-8205-790-1-L8}) mounted on the Palomar
Observatory 1.5-m telescope. These observations were taken from 13
August 2013 to 25 May 2014 (UT). Robo-AO
has a field of view of 44 arcsec $\times$ 44 arcsec at 43.10 mas per
pixel. Typical PSF widths achieved at red-visible wavelengths are in
the range of 0.12 to 0.15 arcsec and companions down to 6 magnitudes
fainter than the primary can be detected
\citep{2014ApJ...791...35L}. We used a Sloan {\it i}'-band
filter \citep{2000AJ....120.1579Y} for stars with $V<13$ and a
long-pass filter cutting on at 600 nm (hereafter LP600) for fainter
stars. Observations consisted of a sequence of full-frame-transfer
detector readouts of the electron-multiplying CCD camera at the
maximum rate of 8.6 Hz for a total of 90 sec of integration time. The
individual images were corrected for detector bias and flat-fielding
effects before being combined through post-facto shift-and-add
processing that used the target as the tip-tilt reference star with 100\% frame
selection.

\section{FALSE POSITIVES} \label{sec-fp}

The NUV-luminous M dwarfs in our sample (i.e., blue points in
Figure~\ref{fig-uv} with spectral subtypes $\leq{\rm M3}$; see
Sections~\ref{sec-bright} \& \ref{sec-latem}) are
presumably active due to their youth. However, early-M dwarfs can
appear NUV-luminous for reasons other than stellar youth. These ``false
positive" (FP) systems include: single early-M dwarfs with unresolved
background NUV sources within the $\sim$5 arcsec beam of \galex{}; unresolved
older binaries where one component is an early-M dwarf and the other
component has persistent NUV emission (e.g.,
white dwarf or late-M dwarf); and short-period
(${\rm P} < 10$ days) tidally interacting binaries that induce ongoing
activity in each other through spin-orbit synchronization.

We identified these FPs in our NUV-luminous sample using two
approaches. First, we used literature searches
(i.e., SIMBAD queries followed with checks in the literature) to
identify known FPs (Section~\ref{sec-lit}). Second, we identified new FP
systems using four detection methods (described in Section
\ref{sec-newfp} and summarized in Table~\ref{tab-methods})
for which we also determined observational completenesses ($C$). We
used the results of the second approach in a maximum likelihood scheme
to estimate the overall FP rate in our NUV-luminous sample
(Section~\ref{sec-fpr}). This allowed us to clean our sample of all
identified FPs and then  statistically correct the remaining
sample for FPs when constructing the NUVLF (Section~\ref{sec-analysis}).

\capstartfalse
\begin{deluxetable}{lrrr}
\small
\centering
\tablewidth{0pt}
\tablecaption{Methods for Detecting False Positives \label{tab-methods}}
\tablecolumns{4}
\tablehead{
   \colhead{Method}
 & \colhead{No. Obs.\textsuperscript{a}}
 & \colhead{No. FPs \textsuperscript{b}}
 & \colhead{Comp. (\%)\textsuperscript{c}}
 }
\startdata
Robo-AO & 193  & 26 & 94 \\
H$\alpha$ Emission & 562  & 37 & 100 \\
H$\alpha$ Centroids & 242 & 25 & 96 \\
SuperWASP  & 312  & 15 & 83 \\
\enddata
\tablenotetext{a}{Number of stars observed using this FP detection
  method. \textsuperscript{b}Number
  of FPs found. \textsuperscript{c}Observational completeness.}
\end{deluxetable}
\capstartfalse

\subsection{SIMBAD Searches} \label{sec-lit}

We queried SIMBAD for our entire sample of NUV-luminous stars in order
to identify any known FPs. We first searched for tight binaries including
spectroscopic binaries (SBs), eclipsing binaries (EBs), and RS Canum
Venaticorum (RS CVn) binaries. We followed up these candidates in the
literature to confirm that they had orbital periods $<10$ days, at which point tidal
interactions between companions likely result in synchronized orbits and
therefore enhanced, persistent NUV emission beyond stellar youth (e.g.,
see \citealt{2006ApJ...653..621M} and references therein). We then
searched for close binaries with separations $<5$ arcsec (i.e., unresolved by
\galex{}) and white dwarf or late-M secondary components, which also emit
persistent NUV emission at older ages. We did not remove
systems containing secondary M dwarf components with unknown spectral
subtypes. We also searched for sources in our sample with background NUV objects
within 5 arcsec, then inspected each of them individually to confirm that they
were not double entries in the SIMBAD database. 

We found 7 EBs, 7 RS CVn systems, and 8 SBs in our sample. We also found
18 close binaries with late-M or white dwarf components within 5
arcsec. There were 9 systems with background sources that were the
probable source of NUV emission, rather than the M dwarf. These
background FPs included a variety of sources: one contact EB,
two SBs, one RS CVn star, two late-M dwarfs, one white
dwarf, and two flare stars.

\subsection{Detecting New False Positives} \label{sec-newfp}

\subsubsection{Robo-AO: Late-M Companions} \label{sec-cont}

We searched our 193 Robo-AO images (see Section \ref{sec-robo}) for
close binaries whose secondary components may be causing
FP NUV emission. We first searched for binaries using a
principal component analysis (to flag elongated sources) and a Gaussian
source finder (to flag multiple sources). We followed up these
candidate binaries with manual (by-eye) checks to confirm the
existence of clearly resolved binaries. We then
developed a method that utilizes binary contrast ratios to identify systems
where the secondary component is likely to be a late-M dwarf with
elevated NUV emission (see Section~\ref{sec-latem}).

\capstartfalse
\begin{figure}
\begin{centering}
\includegraphics[width=8.5cm]{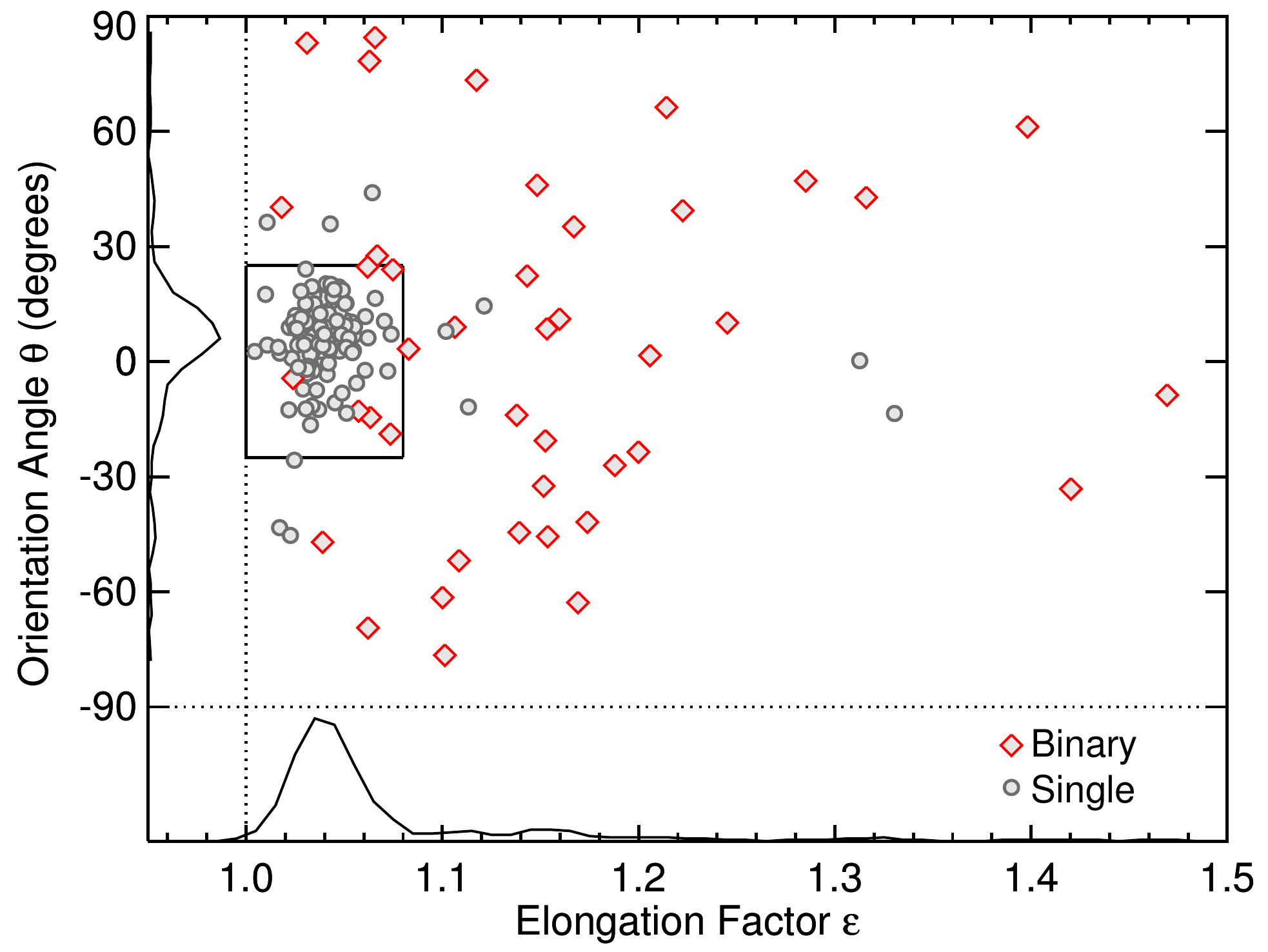}
\caption{\small Confirmed binaries (red diamonds) and single stars
  (gray circles) from our Robo-AO binary search. We used a
  principal component analysis to flag elongated sources outside the
  black box, and a Gaussian source finder to flag multiple sources
  inside the black box. These two methods were required to identify
  candidate binary systems due to the preferential elongation
  ($\varepsilon\sim1.05$) and positive tilt ($\theta\sim10^{\circ}$)
  seen in our dataset, as illustrated by the histograms. All candidate binaries
  were followed up with by-eye checks to confirm those as clearly
  resolved binaries for further analysis.}
\label{fig-ax}
\end{centering}
\end{figure}
\capstartfalse

For the principal component analysis we calculated for each star an elongation
factor, $\varepsilon$ (the ratio between the longer principal axis and
the shorter principal axis), and an orientation angle, $\theta$ (the angle
between the longer principal axis and the vertical image axis; the vertical
axis of Robo-AO images is 23.9$^{\circ}$
right of North). Only sources that were $>10\sigma$ above the noise
were considered, and the noise was calculated using an outlier-resistant
estimate of the dispersion in an area of empty sky around the
source. In theory, single stars should have uniformly distributed
$\theta$ with $\varepsilon=1$, while
binary stars should also have uniformly distributed $\theta$ but
$\varepsilon>1$ due to a companion skewing the otherwise symmetric
distribution. However, Figure~\ref{fig-ax} illustrates that point
sources in our dataset tend to have slight
elongation ($\varepsilon\sim1.05$) and positive tilt
($\theta\sim10^{\circ}$). To account for this in the principal component analysis we
ignored sources with these systematic effects (i.e., all stars inside
the black box in Figure~\ref{fig-ax}) and flagged all other sources as
potential binaries due to their significant elongation. The Gaussian source finder was used
to search the remaining parameter space (i.e., inside the black box in
Figure~\ref{fig-ax}) by looking for positive
brightness perturbations that were $>3\sigma$ above the noise. We
flagged all images with multiple positive brightness perturbations as
potential binaries. We then used by-eye checks on all candidate
binaries to identify only the clearly resolved systems for further
analysis. These binaries and their calculated separations are listed
in Table~\ref{tab-robo}.

Despite the high resolution of Robo-AO, even the closest
binaries resolved by this instrument are too far apart (i.e., several
AU) to be tidally locked. However, Robo-AO can easily resolve binaries with
separations $<5$ arcsec and thus unresolved with \galex{}. Therefore
contrast ratios derived from Robo-AO images can be used to identify
the binaries in our sample with late-M secondary components likely causing
FP activity. To identify such systems, we utilized the empirical
relation between absolute Sloan {\it i}-band magnitude
(M$_{i}$) and spectral subtype derived in
\cite{2002AJ....123.3409H}. By calculating the difference between
M$_{i}$ at M3.5 and M$_{i}$ at all other spectral subtypes, we
derived the maximum allowable contrast ratio ($\Delta$m$_{lim}$)
as a function of primary spectral subtype such that both
components are early-M dwarfs. We confirmed that $\Delta$m$_{lim}$
was applicable to images taken through either Robo-AO
filter (Sloan {\it i}-band or LP600) by measuring the contrast ratios
of a binary in our sample that was imaged in both bands; the
difference in the measured contrast ratios was only $\sim$0.1 mags.

\capstartfalse
\begin{deluxetable}{lccc}
\vspace{0.7 em}
\tabletypesize{\scriptsize}
\centering
\tablewidth{0pt}
\tablecaption{Robo-AO Binaries \label{tab-robo}}
\tablecolumns{3}
\tablehead{\colhead{Star} & \colhead{$\rho$ (arcsec)} & \colhead{$\sigma_{\rho}$ (arcsec)\textsuperscript{a}} & \colhead{Contrast (mag)}}
\startdata
PM\_I00234+2418  & 2.20  & 0.01  & 3.13    \\
PM\_I00235+0947S & 3.68  & 0.01  & 0.27    \\
PM\_I00235+2014  & 1.69  & 0.01  & 1.11    \\
PM\_I00505+2449S & 1.00  & 0.01  & 0.81    \\
PM\_I00531+4829  & 1.26  & 0.01  & 1.59    \\
PM\_I00574+3736  & 4.58  & 0.02  & 5.36    \\
PM\_I01133+5855  & 2.07  & 0.01  & 3.44    \\
PM\_I01146+2057  & 1.42  & 0.01  & 2.12    \\
PM\_I01376+1835  & 1.69  & 0.01  & 0.09    \\
PM\_I01410+5308E & 4.01  & 0.01  & 1.21    \\
PM\_I01480+4652  & 1.13  & 0.01  & 1.53    \\
PM\_I01491+0624  & 0.68  & 0.01  & 0.99    \\
PM\_I02024+1034  & 0.88  & 0.01  & 0.36    \\
PM\_I02208+3320  & 1.46  & 0.01  & 1.88    \\
PM\_I02347+1251  & 1.20  & 0.01  & 1.37    \\
PM\_I02408+4452  & 6.03  & 0.02  & 5.04    \\
PM\_I02560+1220  & 0.90  & 0.01  & 0.75    \\
PM\_I03053+2131  & 0.62  & 0.01  & 0.57    \\
PM\_I04284+1741  & 1.68  & 0.01  & 1.82    \\
PM\_I04310+3647  & 0.80  & 0.01  & 0.19    \\
PM\_I04333+2359  & 0.74  & 0.01  & 0.61    \\
PM\_I04453+1334  & 5.58  & 0.02  & 4.09    \\
PM\_I04499+2341E & 2.39  & 0.01  & 0.83    \\
PM\_I04540+2200  & 3.85  & 0.02  & 3.81    \\
PM\_I05228+2016  & 3.61  & 0.01  & 1.78    \\
PM\_I05341+4732  & 2.47  & 0.01  & 1.01    \\
PM\_I06088+4257  & 1.25  & 0.01  & 0.75    \\
PM\_I06212+4414  & 1.37  & 0.02  & 3.18    \\
PM\_I06268+4202  & 0.85  & 0.01  & 1.25    \\
PM\_I11505+2903S & 0.49  & 0.01  & 0.61    \\
PM\_I17038+3211  & 1.32  & 0.02  & 1.84    \\
PM\_I20514+3104  & 1.41  & 0.01  & 1.17    \\
PM\_I21010+2615  & 0.44  & 0.01  & 0.18    \\
PM\_I21221+2255  & 5.20  & 0.02  & 4.81    \\
PM\_I21410+3504  & 4.60  & 0.01  & 0.30    \\
PM\_I22006+2715  & 5.43  & 0.03  & 5.14    \\
PM\_I22234+3227  & 1.34  & 0.01  & 0.51    \\
PM\_I23045+4014  & 0.83  & 0.01  & 0.62    \\
PM\_I23063+1236  & 0.42  & 0.01  & 0.36    \\
PM\_I23300+1643  & 0.94  & 0.01  & 0.22    \\
PM\_I23318+1956W & 5.40  & 0.01  & 1.75    \\
PM\_I23450+1458  & 1.17  & 0.01  & 0.27    \\
PM\_I23535+1206S & 5.72  & 0.01  & 0.80    \\
PM\_I23578+3837  & 0.52  & 0.01  & 1.65    \\
\enddata
\tablenotetext{a}{Separation errors were calculated from uncertainties on the binary centroid positions, which were produced when using the IRAF {\it phot} routine to measure contrast ratios (see Section~\ref{sec-cont}).}
\vspace{1 em}
\end{deluxetable}
\capstartfalse

\begin{figure}
\begin{centering}
\includegraphics[width=8.5cm]{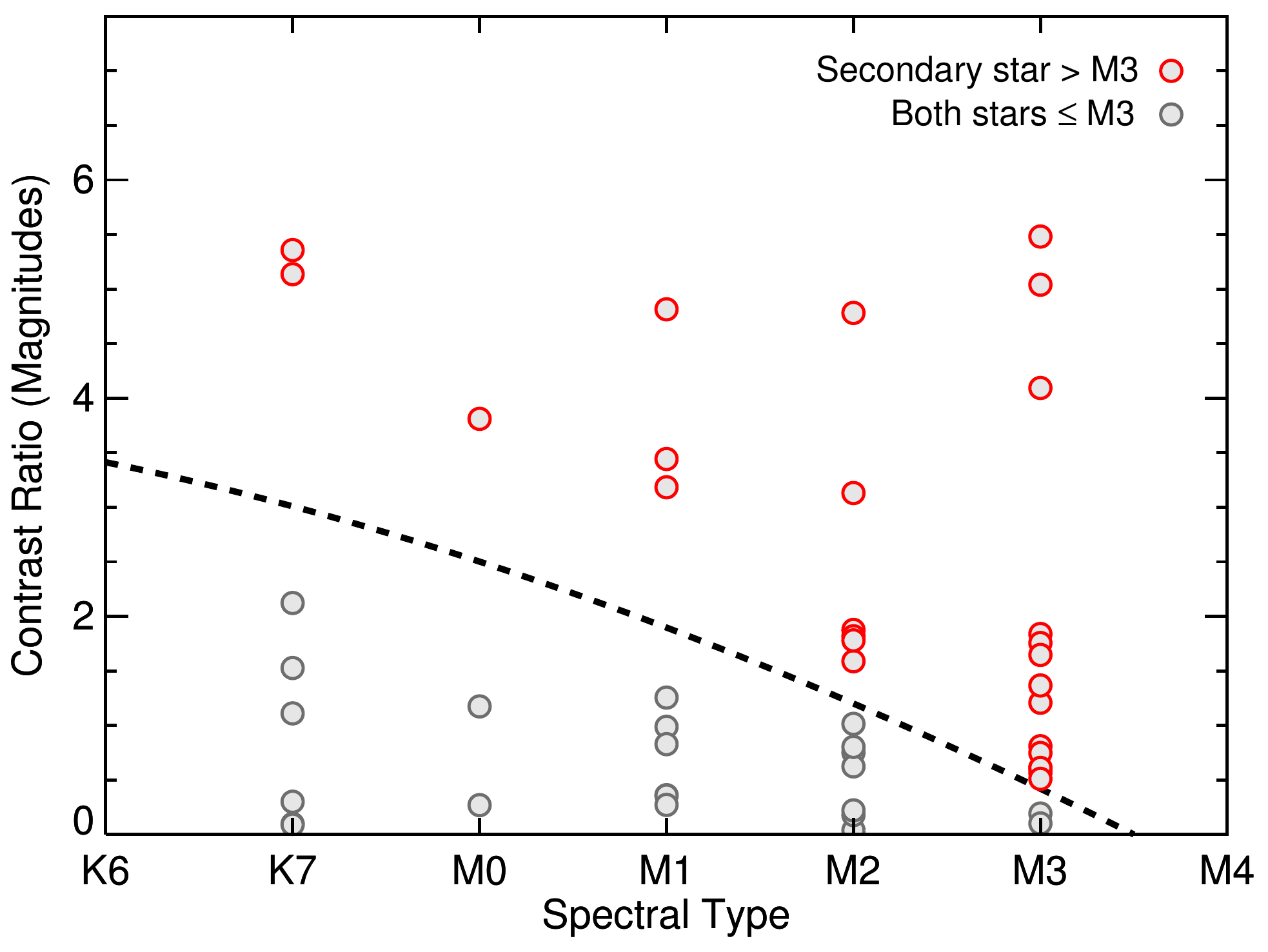}
\caption{\small Contrast ratios for systems identified by Robo-AO
  as having secondary components within 5 arcsec. The dashed line
  shows the maximum contrast ratio as a function of primary
  spectral subtype for systems where both
  stars are early-M dwarfs. Systems above the dashed line
  are predicted to have companions with spectral subtypes later than M3.}
\label{fig-cont}
\end{centering}
\end{figure}

For each Robo-AO binary, we measured the contrast ratio by performing
aperture photometry on each companion using the IRAF {\it phot}
routine and a 20-pixel circular aperture. This aperture size was
based on when the curve of growth for a typical single star reached an
asymptotic value. Closer binaries required
smaller aperture radii (5--15 pixels) to avoid contamination from
companions. For sky subtraction we used median-combined, manually
sampled patches of nearby empty sky around each
system to ensure proper sky measurements. Figure~\ref{fig-cont} shows
measured contrast ratios for all our Robo-AO binaries with separations $\lesssim5$
arcsec. The black dashed line indicates our calculated $\Delta$m$_{
  lim}$ as a function of primary spectral subtype. We found 26
binaries with primary spectral subtypes $\leq {\rm M3}$ but
secondary components likely to be $> {\rm M3}$, making them potential FPs.

The observational completeness of our Robo-AO FP search was
limited by two factors: (i) the maximum contrast ratio that Robo-AO can detect
and (ii) the probability that a late-M companion was unresolved
because its projected distance from the primary at the time of
observation was too small. To estimate the completeness due to (i), we
used the \cite{2007AJ....133..439C} $J$-band luminosity function,
$n$(M$_{J}$), to calculate the number of M dwarf companions expected
to have spectral subtypes $\ge {\rm M3.5}$ but earlier than the limit
imposed by the maximum contrast ratio that Robo-AO can detect,
$\Delta$m$_{\rm max}$. We then compared that number to the total number of M
dwarf companions expected to have spectral subtypes $\ge {\rm M3.5}$:

\begin{equation}\label{eq-crobo}
  C =\frac{\sum_{i=1}^{binaries} \int_{\rm M_{\it J, 3.5}}^{\rm M_{\it
        J, SpT_i}+\Delta m_{{\rm max}, {\it i}}} n(\rm M_{\it J})dM_{\it J} }{\sum_{i=1}^{binaries} \int_{\rm M_{\it J, 3.5}}^{M_{J, 9.0}} n(\rm M_{\it J})dM_{\it J}} .
\end{equation}

We adopted $\Delta$m$_{\rm max}=6.0$, which is the maximum contrast ratio that
Robo-AO can achieve at high ``contrast-performance'' (i.e., PSF sizes
$>0.15$ arcsec) for separations $>1.0$ arcsec in the Sloan {\it i}-band
\citep[see Figure 5 in][]{2014ApJ...791...35L}. The
median PSF size and separation for the binaries in our Robo-AO sample were
$\sim$0.17 arcsec and $\sim$1.41 arcsec, respectively,
justifying this choice of $\Delta$m$_{\rm max}$. We used relations
from \cite{2002AJ....123.3409H} to obtain $i-J$ color as a function of
M$_{J}$, which allowed us to translate our Robo-AO contrast ratios
from $i$- to $J$-band for use in Equation~\ref{eq-crobo}. We
calculated each primary's M$_{J}$ using its $V-J$ color
according to Equation 22 from \cite{2013AJ....145..102L}. Evaluating
Equation~\ref{eq-crobo} gave a completeness factor of $\sim$0.97.

To estimate completeness due to (ii), we compared the median PSF
size in our Robo-AO images ($\sim$0.17 arcsec) to the \galex{} PSF size
($\sim$5.0 arcsec) and assumed a uniform source distribution on the
sky \citep[equivalent to an isotropic, uniform distribution of orbits
with a log orbital period;][]{1991A&A...248..485D}. This gave a
$\sim$$3\%$ probability of an unresolved companion and thus a
$\sim$0.97 completeness factor. The product the completenesses from (i)
and (ii)  gave a final observational completeness of $C\sim0.94$ for our
Robo-AO FP search.

\capstartfalse
\capstartfalse

\subsubsection{Optical Spectra: Missing H$\alpha$ Emission} \label{sec-halpha}

Emission in the Balmer $\alpha$ line of hydrogen (\ha{}; $\lambda
\approx 6563$\AA) is another indicator of stellar activity and is
strongly correlated with NUV emission
\citep[e.g.,][]{2013AJ....145..102L}.  FPs resulting from
\galex{} source confusion can therefore potentially be identified by
the absence of H$\alpha$ emission despite significant amounts of NUV
emission (i.e., NUV emission at least $2.5\sigma$ above the basal
level; see Section~\ref{sec-bright}). This is because sources that are
unresolved by \galex{}'s $\sim$5 arcsec beam can be resolved by
spectroscopy with $\sim$1 arcsec resolution. We therefore computed
\ha{} equivalent widths (EW$_{\rm H \alpha}$) for the 2128 stars in
our sample with medium-resolution optical spectra
(see Section \ref{sec-spectra}). 

To calculate EW$_{\rm H \alpha}$ we first shifted each spectrum to
its rest frame by applying wavelength offsets found by matching
our observed spectra to PHOENIX model atmospheres. We used the
BT-SETTL version of the PHOENIX atmospheric model code (\citealt{2013MSAIS..24..128A};
\citealt{2013A&A...556A..15R}) with the CIFIST grid and \citet{2011SoPh..268..255C}
abundances for the Sun. Following \cite{2013AJ....145..102L}, we
measured the flux within a 14\AA-wide spectral region
(6557.61--6571.61\AA{}, in air) relative to pseudo-continuum regions
(6500--6550\AA{} and 6575--6625\AA, in air).  Errors were calculated
using a Monte Carlo method that assumed Gaussian-distributed noise and
random wavelength calibration errors of 0.5\AA{}. As noted by
\cite{2013AJ....145..102L}, this choice of continuum region
systematically underestimates EW$_{\rm H \alpha}$ values due to
differences between our pseudo-continuum and the true spectral
continuum. We therefore applied a small offset (0.3\AA) so that
stars with basal NUV emission had a mean EW$_{\rm H
  \alpha}\approx0$. This did not affect our FP analysis
as the offset was applied to the entire population.

\begin{figure}
\begin{centering}
\includegraphics[width=8.5cm]{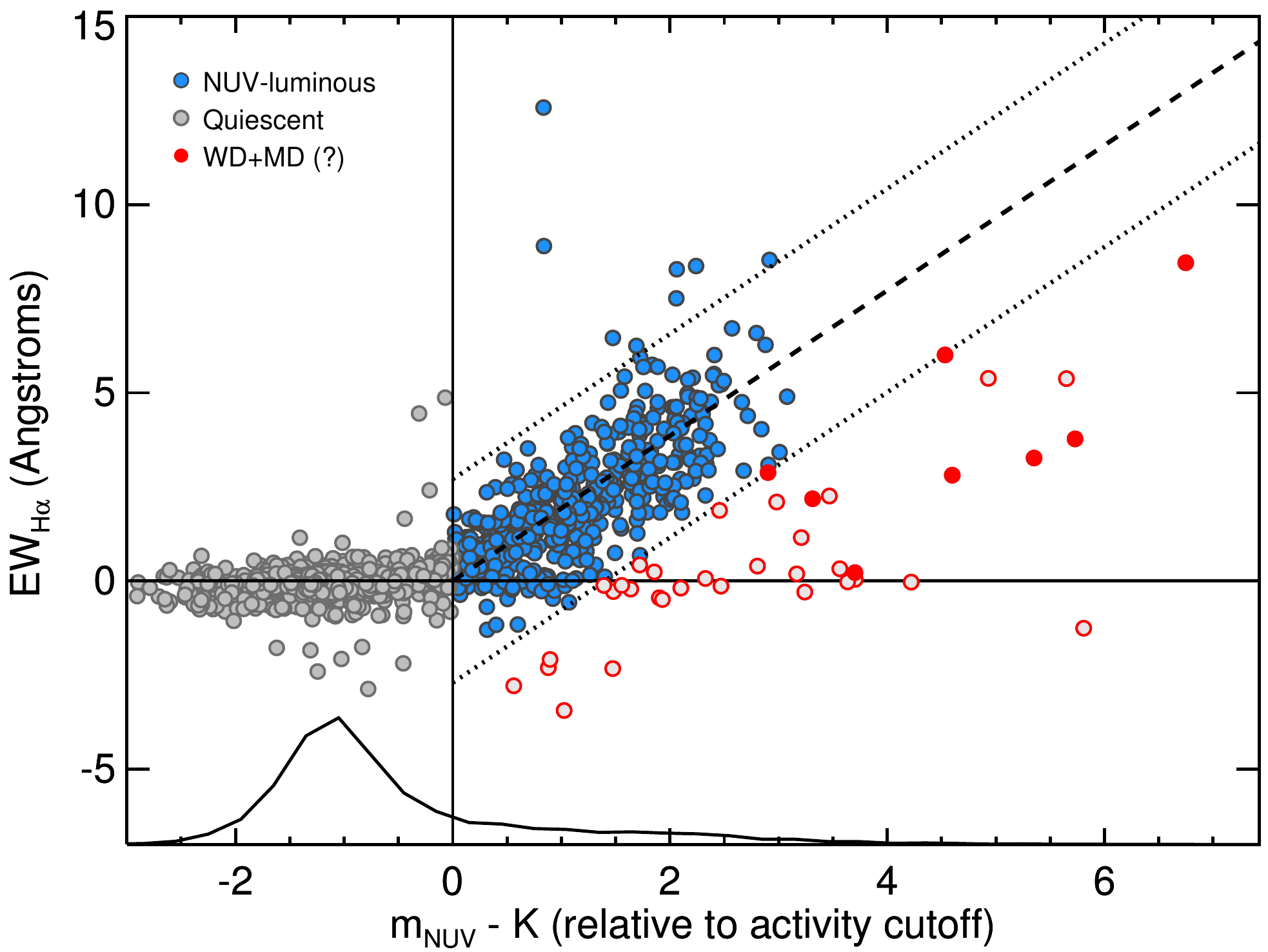}
\caption{\small EW$_{\rm H \alpha}$ as a function of our selection
  cutoff for NUV-luminous stars (see Section \ref{sec-bright}). Blue and gray points are
  the same as in Figure~\ref{fig-uv}. The dashed black line is the median fit to
  the NUV-luminous population, representing the expected EW$_{\rm H
    \alpha}$ of a young M dwarf with a given NUV emission level;
  3$\sigma$ boundaries are shown by dotted black
  lines. Suspect FPs, outlined in red, reside at least 3$\sigma$
  below their expected EW$_{\rm H \alpha}$
  value. Filled red circles indicate known or candidate
  WD+MD systems (see Section \ref{sec-other}).}
\label{fig-halpha}
\end{centering}
\end{figure}

Figure~\ref{fig-halpha} shows our measured EW$_{\rm H \alpha}$ values
as a function of our selection cutoff for NUV-luminous stars (see Section
\ref{sec-bright}). As expected, the stars with basal NUV emission
(gray points, corresponding to those in Figure~\ref{fig-uv}) are
located in a band with negligible EW$_{\rm H \alpha}$. The
NUV-luminous stars (blue points, corresponding to those in
Figure~\ref{fig-uv}) form a distinct locus where increasing NUV
emission corresponds to increasing EW$_{\rm H \alpha}$ values. However, a
subset of NUV-luminous stars have lower-than-expected EW$_{\rm H
  \alpha}$ values (i.e., they lie significantly below the NUV-luminous locus)
and are presumably FPs due to \galex{} source confusion. We fit a line to
median values of EW$_{\rm H\alpha}$ vs. relative NUV emission for the
NUV-luminous population (black dashed line in Figure~\ref{fig-halpha}) to
determine the expected EW$_{\rm H \alpha}$ value for a young M dwarf with a
given NUV emission level. We identified 37 stars with EW$_{\rm H
  \alpha}$ values more than $3\sigma$ {\it below} the
expected value, making them likely FPs.

\capstartfalse
\capstartfalse

Interestingly, all known WD+MD binaries in our sample as well as the new candidate
WD+MD systems identified in this work (see Section~\ref{sec-other})
were flagged as FPs using this detection method. As shown in
Figure~\ref{fig-halpha}, these WD+MD pairs have
anomalously high NUV emission relative to their EW$_{\rm H \alpha}$
values, as expected (see Section~\ref{sec-other}). It is also
important to note the distinct lack of stars with high EW$_{\rm H
  \alpha}$ values but basal NUV emission levels (upper left quadrant
in Figure~\ref{fig-halpha}); because the NUV
and H$\alpha$ measurements were taken at different epochs, this
may illustrate consistent levels of activity for the vast majority of
stars in our sample. This would suggest that our sample contains
very few flare stars and thus a low probability of
misidentified FPs due to stellar variability. Of course the same
reasoning could be applied to the significant population of sources with high NUV
emission but low EW$_{\rm H \alpha}$ values to argue for evidence of flaring; however,
as discussed above, this population also contains FPs, which
complicates the interpretation.  

We estimated the observational completeness for this FP detection
method using an injection and recovery method. We replaced
the region of H$\alpha$ emission in each star's spectrum with the median of the
surrounding continuum flux, then injected a synthetic H$\alpha$
signal with an equivalent width equal to the expected value for that star's
NUV emission level (using the black dashed line in
Figure~\ref{fig-halpha} as a guide). We also added random Gaussian noise scaled
to the noise in the surrounding continuum regions. We re-measured the
EW$_{\rm H \alpha}$ values using the same method as above, repeating 100
times and taking the average of the resulting equivalent widths for each
star. We then checked whether this EW$_{\rm H \alpha}$ value was
within 3$\sigma$ of the expected value. In all cases the signal was
recovered, implying $C\sim1.0$ for our EW$_{\rm H \alpha}$ FP detection method.

\subsubsection{SNIFS Integral Field Spectra: H$\alpha$-emitting
  Companions} \label{sec-shift}

SNIFS image cubes provide both spatial and wavelength dimensions that
can be used in conjunction to search for FPs. In particular, they can
be used to find binary systems appearing young due to unresolved
late-M companions with persistent activity despite being
old. In such cases, the early-M primary exhibits basal NUV
emission but dominates the continuum signal (making the system appear as
a single early-M dwarf), while the unresolved late-M companion
is the source of the stronger \ha{} emission and presumably also the NUV
flux. This configuration is detectable as a shift between the
centroid of a white-light image versus the centroid of an \ha{}
image. We used this method of detecting FPs in addition to our Robo-AO
image analysis (Section \ref{sec-cont}) because \ha{} traces activity
and we have more SNIFS image cubes than Robo-AO images.

We performed this analysis on the 242 stars in our NUV-luminous sample
with SNIFS image cubes. To create the white-light image we summed the
SNIFS image cube over all wavelengths covered by the spectrum. We
identified the source centroid location by employing a principal
component analysis using only points that were $>10\sigma$ above the
noise. The noise was calculated using an outlier-resistant estimate of
the dispersion around the outer edge of the image; we used the outer
edge due to the small size of SNIFS images (14 pixels$\times$14 pixels). To
create the \ha{} image we summed the SNIFS image cube across only
wavelengths covering the \ha{} spectral line (see Section
\ref{sec-halpha} for our \ha{} line parameters), then subtracted the
continuum and divided by the noise. The continuum was estimated using
the median value of surrounding wavelength regions multiplied by the
number of wavelength elements covered by the spectral line, and the noise was
found by taking the standard deviation of the continuum regions. To identify
the \ha{} centroid location, we again used a
principal component analysis but first applied a mask to consider only
pixels that were also used to calculate the white-light centroid. We
discarded systems lacking any significant \ha{} emission.

\begin{figure}
\begin{centering}
\includegraphics[width=8.5cm]{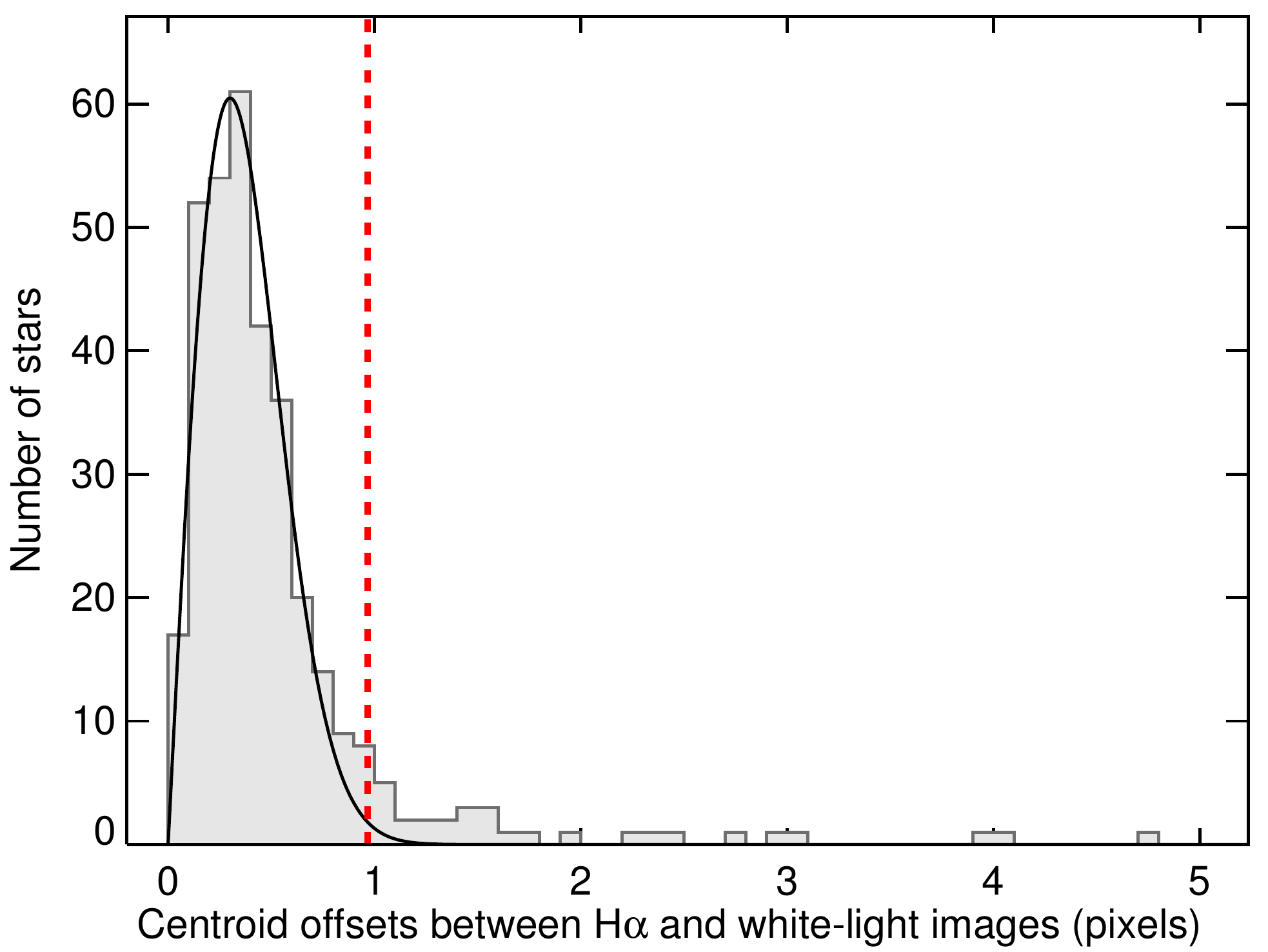}
\caption{\small Centroid offsets between white-light and H$\alpha$
  images for the 242 NUV-luminous sources in our sample with SNIFS
  image cubes. A Rayleigh fit to the distribution (black solid line)
  produced a mean of $\mu=0.37$ and error of $\sigma=0.20$. Systems
  with offsets $>3\sigma$ above the mean (red dashed line) were
  flagged as potential FPs then followed up with by-eye checks to
  confirm the likely existence of an unresolved late-M companion.}
\label{fig-ray}
\end{centering}
\end{figure}

Figure~\ref{fig-ray}  shows the distribution of offsets between the
white-light and H$\alpha$ centroids. We fit a Rayleigh function to the distribution
using the IDL routine {\it mpfit} \citep{2009ASPC..411..251M}, which
returned a mean offset of $\mu\approx0.37$ pixels and a dispersion of
$\sigma\approx0.20$ pixels. We flagged systems with centroid
offsets $>3\sigma$ above the mean as potential FPs, then used by-eye checks to
confirm 25 systems with clear shifts in their image centroids.
We estimated the observational completeness as the ratio of the area over which false
positives could be detected (i.e. the area of the annulus from $r=3\sigma$ to
$r=r_{max}$, where $r_{max}$ is the maximum centroid offset in our
sample) to the total survey area (i.e., the area of a circle with
$r=r_{max}$). This resulted in an observational completeness of
$C\sim0.96$ for our centroid offset detection method.

\capstartfalse
\capstartfalse

\subsubsection{SuperWASP Light Curves: Tidally-Locked Binaries} \label{sec-fast}

Time-series photometry from the SuperWASP exoplanet survey
\citep{2006PASP..118.1407P} can be repurposed to identify very short-period,
interacting stellar binaries \citep[e.g.,][]{2011A&A...528A..90N}.  These
systems of tidally-locked, synchronously rotating stars remain
magnetically active and rapidly rotating due to the transfer of
angular momentum from their orbits to their spins. They exhibit
photometric variability because of transits and/or fixed
patterns of spots established by the interacting magnetic fields of
the companions. Even though a young, single M dwarf can also have a surplus
of star spots due to elevated activity, these spots tend to be
uniformly distributed across the surface, which likely dampens any
induced light curve variability \citep[see][and references
  therein]{2011MNRAS.412.1599B}. Moreover spots on single stars tend to
migrate, causing the phases of their light curve signals to change over time.  

We therefore cross-referenced our NUV-luminous sample with the
SuperWASP database, which is available
online.\footnote{http://exoplanetarchive.ipac.caltech.edu}
We used a 3 arcsec search radius and only considered light curves with
more than 1,000 data points due to the limited photometric
precision of SuperWASP. This resulted in a sample of 312 NUV-luminous sources with
SuperWASP light curves. We inspected each of these light curves for
stellar variability by computing their Lomb-Scargle periodogram
\citep{1987BAAS...19Q.829S}, which estimates a frequency spectrum
based on a least-squares fit to sinusoids.  We
only considered signals with a false-alarm probability of $<0.1$\%.
We ignored periods within 2\% of one day and fractions thereof (1/4,
1/3, 1/5, etc.) as those are likely artifacts due to observing
schedules \citep{2014MNRAS.437.3133G}. We used a stricter
10\% filter around periods with stronger systematics (0.5, 1.0, 1.5,
and 2.0 days).

\capstartfalse
\begin{figure}
\begin{centering}
\includegraphics[width=8.5cm]{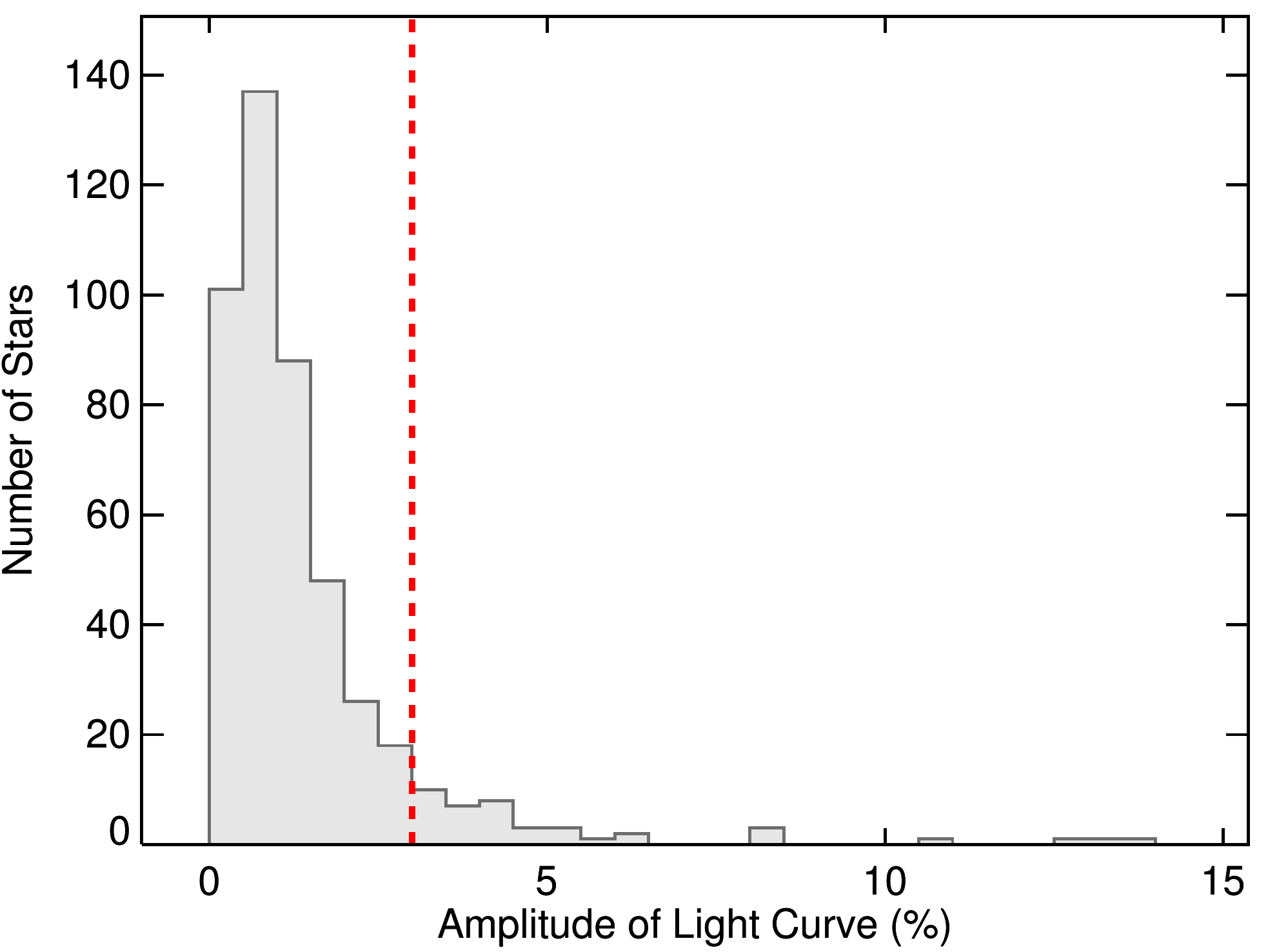}
\caption{\small Amplitudes of all SuperWASP light curves available
  for our NUV-luminous sources. The dashed line shows the
  amplitude cutoff above which sources were considered candidate
  tidally interacting binaries (if they also fulfilled our period
  criteria; see Section \ref{sec-fast}).}
\label{fig-wasp}
\end{centering}
\vspace{0.4 em}
\end{figure}
\capstartfalse

To identify candidate tidally interacting binaries we selected light
curves that featured large amplitude variations ($>2.6$\%) at short
periods ($<10$ days) with phases that appeared perfectly Keplerian
(i.e., stable over many cycles). The period criterion was based on the
orbital period at which tidal interactions between companions begin to
synchronize orbits and therefore enhance stellar activity \cite[see][and references
therein]{2006ApJ...653..621M}. The amplitude criterion was obtained from running the
Lomb-Scargle periodogram on all known eclipsing binaries in LG11 with
SuperWASP light curves, then taking the minimum amplitude recovered as
our lower limit. 

Figure~\ref{fig-wasp} shows this limit in the context of
all SuperWASP light curve amplitudes derived from our sample. From the
preliminary list of candidate FPs generated by applying the above criteria,
we used by-eye checks to identify 15 light curves with clear sinusoidal signals
consistent with those of known tidally locked binaries. We then divided each of these light
curves into halves and re-ran the analysis on both sections to check
that the period remain unchanged, indicating variability due to
regular eclipses rather than varying star spot patterns.

We estimated observational completeness using a method of injection and recovery
of artificial sinusoidal signals. We randomly selected 100 SuperWASP
light curves from our sample and randomized the fluxes of the data
points. We then injected signals with randomly selected periods of
0.1--10 days (i.e., our FP period search criteria) and amplitudes of 2.6--6.7\%
(i.e., our FP amplitude search criteria limited by the maximum amplitude found
in our sample). We then re-performed our Lomb-Scargle periodogram search,
using the same period filters and false-alarm probability as before,
to test whether we would have recovered the injected signals as
candidate FPs. This produced an observational completeness of $C
\sim0.83$ for our SuperWASP FP detection method.

\medskip
\subsection{Derivation of an Overall False Positive Rate} \label{sec-fpr}

Construction of an accurate NUVLF requires the identification and
consideration of FPs, i.e. systems that appear NUV-luminous for reasons
other than stellar youth. Our approach was to (i) estimate the overall
FP rate among our NUV-luminous sample based on FP detection methods for which
we could determine observational completenesses (Section \ref{sec-newfp}); (ii) remove
FPs identified using these FP detection methods as well as known FPs from
the literature (Section \ref{sec-lit}); and then (iii) use our
derived overall FP rate to statistically correct the remaining
NUV-luminous stars in our sample not yet flagged as FPs
(Section~\ref{sec-lf}).  The last step was necessary to account for the
fact that not all NUV-luminous sources were observed with all FP detection
methods, and also because our FP detection methods cannot each detect all
possible types of FPs (e.g., the Robo-AO method cannot resolve tidally
locked binaries).

We used a maximum likelihood approach to
estimate the overall FP rate in our NUV-luminous sample. When
screening a source with our FP detection methods in Section
\ref{sec-newfp}, the source is either identified as a FP or it is
not. We can describe the likelihood of these two possible outcomes,
$f(p)$ and $g(p$), respectively, in terms of the overall
FP rate ($p$) and the completenesses of the different
FP detection methods that were applied to the source ($C_{j}$):

\begin{equation}\label{eq-mlnfp}
g(p) = (1-p) + p \prod_{j}^{missed}  (1-C_{j})
\end{equation}

and 

\begin{equation}\label{eq-mlfp}
f(p) = p \big[1-\prod_{j}^{missed}(1-C_{j})\big] .
\end{equation}

Equation~\ref{eq-mlnfp} is the probability that a source was {\it not}
identified as a FP: the first term is the probability that the
source was not a FP, while the second term is the probability that the
source was a FP but that it was missed by all the FP detection methods
that were applied to it. Equation~\ref{eq-mlfp}
is the probability that a source was identified as a FP: the term in the brackets
is the probability that the source was identified as a FP by the
detection methods that were applied to it, and this term is then
multiplied by the actual false-positive rate. This
approach assumes that there are no ``false negatives" (i.e., it
assumes that our FP detect methods cannot wrongly classify a source as a
FP). It also assumes that a source is a FP if just one of our methods
detected it as a FP. We found the value of $p$ that maximizes the likelihood: 

\capstartfalse
\begin{figure}
\begin{centering}
\includegraphics[width=8.5cm]{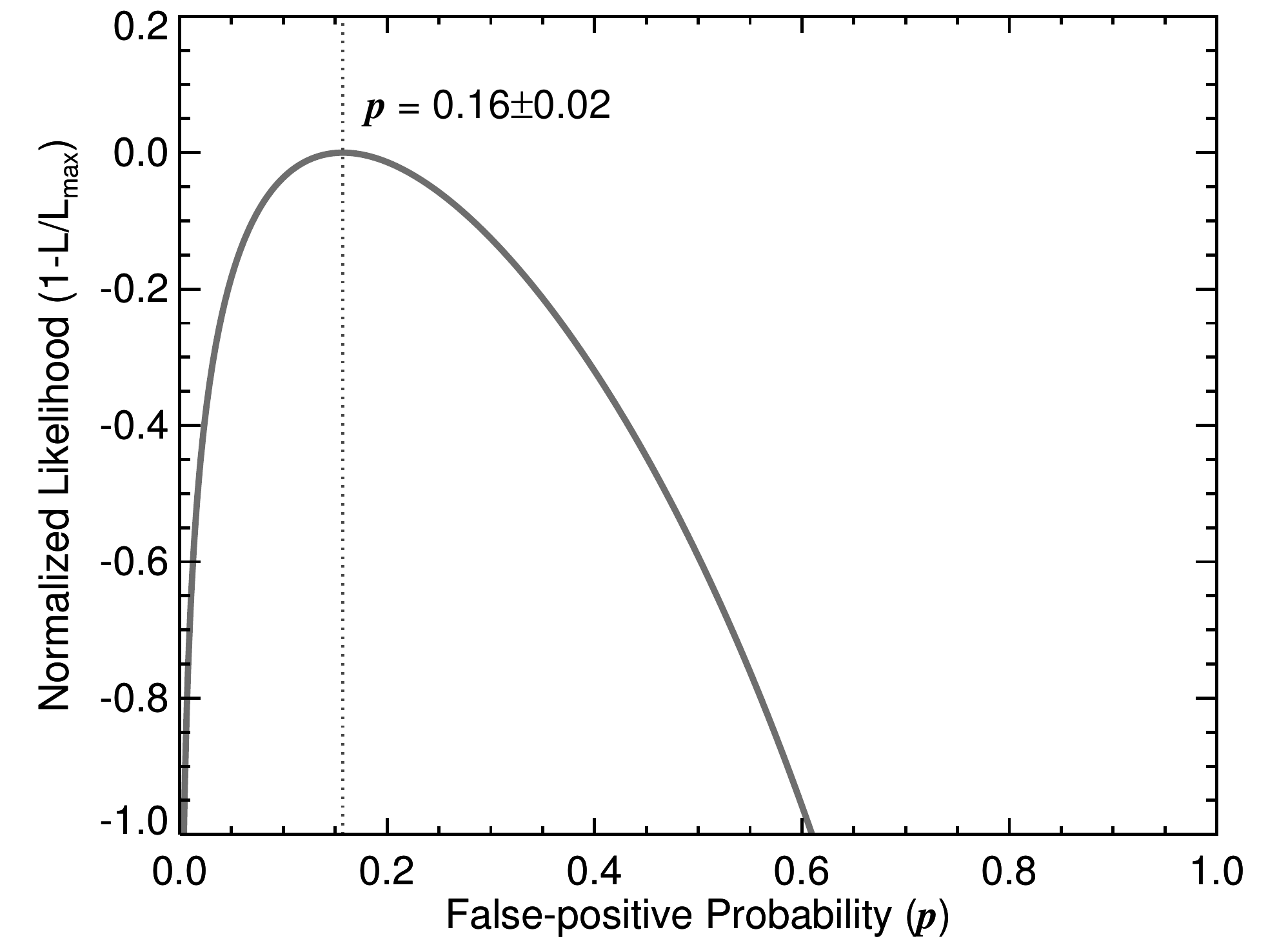}
\caption{\small The maximum likelihood (see
  Equation~\ref{eq-ml2}) used to
  determine the most probable value of the overall FP rate, $p$, in our
  NUV-luminous sample of early-M dwarfs.}
\label{fig-ml}
\end{centering}
\end{figure}
\capstartfalse

\begin{equation}\label{eq-ml2}
\ln \mathscr{L} = N_{\rm FP} \ln p + \sum_{k}^{\rm NFP} \ln \big[ 1- p \big(1 - \prod_{m} (1-C_{m}) \big) \big] ,
\end{equation}

\noindent where $N_{\rm FP}$ is the total number of FPs found using
our FP detection methods, and $C_{m}$ is the observational
completeness of each FP detection method applied to a given
star that was {\it not} found to be a FP (NFP). Solutions to Equation~\ref{eq-ml2} for
all possible values of $p$ are shown in Figure~\ref{fig-ml}, indicating a most
likely FP rate of $p\sim0.16\pm0.02$.  We estimated the errors by
fitting an inverted parabola around the peak in the log likelihood, then using the
uncertainty on the curvature of the fit (i.e., $\sigma_{x}=1/\sqrt{-2c}$
for a parabola described by $y=a+bx+cx^2$). 

\bigskip

\section{THE NUV LUMINOSITY FUNCTION} \label{sec-analysis}

\subsection{Fractional NUV Luminosity} \label{sec-frac}

We defined a fractional NUV luminosity to describe
the distribution of NUV luminosities with respect to a basal value:

\begin{equation}\label{eq-frac}
R'_{\rm NUV} = \frac{\rm L_{\rm NUV}-\rm L_{\rm basal}}{\rm L_{bol}} ,
\end{equation}

\noindent where L$_{\rm NUV}$ is the total NUV luminosity, L$_{\rm basal}$
is the basal NUV luminosity, and L$_{\rm bol}$ is the
bolometric luminosity. To obtain these parameters from the
observed quantities shown in Figure~\ref{fig-uv}, we re-arranged the
color relation $m_{\rm NUV}-K_{\rm S}=M_{\rm NUV}-{\rm M}_{K_{\rm S}}$ to obtain
an expression for absolute NUV magnitude: ${\rm M}_{\rm NUV}={\rm
  M}_{K_{\rm S}}+(m_{\rm NUV}-K_{\rm S})$. We calculated M$_{K_{\rm S}}$
by translating $V-J$ color into M$_{J}$ using Equation 22 from
\cite{2013AJ....145..102L} then converting M$_{J}$ to M$_{K_{\rm S}}$ using
$J-K_{\rm S}=0.854$ ($J-K_{\rm S}$ color varies little for early-M
dwarfs; here we used the median of our sample). To translate M$_{\rm
  NUV}$ into L$_{\rm NUV}$ we used the \galex{} zero points\footnote{ \scriptsize
  http://galexgi.gsfc.nasa.gov/docs/galex/FAQ/counts\textunderscore
  background.html} to convert magnitude into flux density (ergs$^{-1}$
s$^{-1}$ cm$^{-2}$ \AA$^{-1}$) then multiplied by the effective
\galex{} NUV bandwidth ($\Delta\lambda =732$ \AA) to get the NUV flux,
F$_{\rm NUV}$. We then input F$_{\rm NUV}$ into L$_{\rm NUV}=4\pi
d^{2}\rm F_{\rm NUV}$ where $d=10$ pc (in accordance with the
definition of absolute magnitude). We also applied this method to the
median fit shown in Figure~\ref{fig-uv} (black dashed line) to obtain L$_{\rm
  base}$ as a function of $V-J$ color. To calculate L$_{\rm bol}$ we used
the standard equation L$_{\rm bol}={\rm L}_{\odot}\times10^{-0.4({\rm M}_{bol}-{\rm M}_{bol, \odot})}$
where M$_{bol, \odot}=4.7554$ magnitudes and L$_{\odot}=3.8270\times10^{33}$ ergs$^{-1}$ s$^{-1}$
\citep{2012ApJ...754L..20M}. To obtain ${\rm M}_{\rm bol}$ we used the
$K_{\rm S}$ bolometric correction from \citet{2001ApJ...548..908L}.

\capstartfalse
\begin{figure*}
\includegraphics[width=17.5cm]{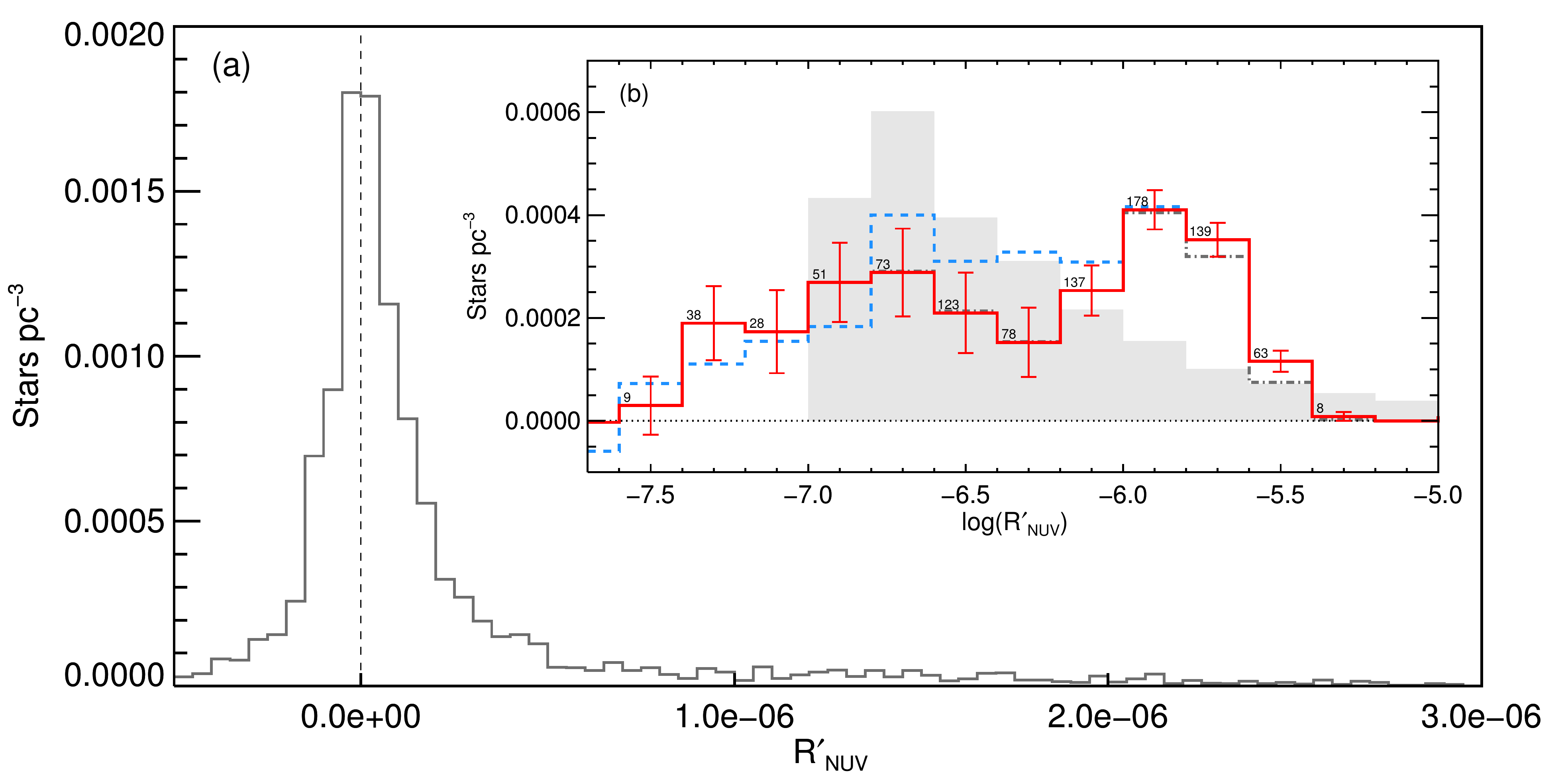}
\caption{\small (a) Our derived NUVLF for young, early-M dwarfs using the $1/V_{
    max}$ method described in Section~\ref{sec-lf}. The peak
  centered on zero consists of stars exhibiting basal NUV emission and the extended
  tail (only partially shown here) consists of NUV-luminous stars that
  are presumably young/active. (b) The residual NUVLF, now shown in
  log space, representing only the young early-M dwarfs. The residual
  distribution in (b) was found by subtracting opposing sides of the distribution
  in (a). The red solid line shows the NUVLF when including upper limits
  (see Section~\ref{sec-avni}), while the blue dashed line shows the NUVLF when
  ignoring upper limits (see Section~\ref{sec-lf}). For the
  residual distribution that includes upper limits, the number of
  stars in each bin is shown for reference and error bars were found
  using bootstrap sampling. The gray dash-dotted line (mostly coincident with
  the red solid line except at the highest $R'_{\rm NUV}$ values)
  shows the NUVLF derived using a varying FP rate (see Section~\ref{sec-varyfp}). The light
  gray filled histogram is the best-fit model assuming a constant
  star-formation rate and an age-activity relation described by a
  two-parameter power law (see Section \ref{sec-age}).}
\label{fig-volume}
\end{figure*}
\capstartfalse

\subsection{The $1/V_{\rm max}$ Method} \label{sec-lf}

The ``$1/V_{\rm max}$" method \citep{1968ApJ...151..393S}
is used to construct luminosity functions by accounting for the bias
of flux-limited surveys toward intrinsically bright sources. This is
done by inversely weighting sources by the volume of space over
which they could have been detected by the survey. Thus
bright sources are assigned smaller weights to correct for being
detectable to larger distances. The luminosity function is then
calculated by summing the weights (rather than number of stars) in each
luminosity bin, giving units of stars pc$^{-3}$. 

To apply the $1/V_{\rm max}$ method to our sample, we calculated for
each star the maximum distance at which it would have been included in LG11 and
also detected by \galex{}. Thus the limiting detection distance of each star
was determined by one of three factors: (i) the
$J$-band magnitude limit ($J<10$) of the LG11 catalog; (ii) the proper
motion limit ($\gtrsim40$ mas in the north, $\gtrsim100$ mas in
the south) of the LG11 catalog; and (iii) the \galex{} sensitivity
limit. Because \galex{} sensitivity varies across the sky due to
varying tile exposure times, we estimated the limiting NUV magnitude
for each star in our sample (see Section~\ref{sec-avni}) to
create a map of limiting NUV magnitude across the sky. Then for each star we
compared its two LG11 limiting detection distances to each
\galex{} limiting detection distance across the sky, recording the smallest
value in each case. The average of the cube of these smallest
detection distances was then used to calculate $V_{\rm max}$ for that star. The star's
contribution to the NUVLF was then determined by its weight, $1/V_{\rm max}$. 

To account for FPs, we first removed all FPs found in the literature
(Section \ref{sec-lit}) or with our FP detection methods (Section
\ref{sec-newfp}). We then statistically accounted for FPs in the remaining stars by
multiplying their $1/V_{\rm max}$ weights by $1-p$ (i.e., the
probability of {\it not} being a FP). For NUV-luminous stars we used
$p=0.16$ (see Section~\ref{sec-fpr}) and for all other stars we used
$p=0$. There were also 92 sources in our sample with limiting detection
distances that were found to be smaller than their actual
distance. This was mostly due to anomalies in survey
sensitivities (e.g., 67 of these sources had declinations $<-20^{\circ}$ where LG11 proper motion limits
can vary). We removed these sources from our sample before
constructing the NUVLF, however only 9 of these sources were members of our
NUV-luminous sample (these are flagged in Table~\ref{tab-params}).

The resulting NUVLF is shown in the main panel of
Figure~\ref{fig-volume}. By construction, the peak at $R'_{\rm NUV}\approx0$ consists
of stars with basal levels of NUV emission (designated by gray points in
Figure~\ref{fig-uv}) while the extended tail toward higher $R'_{\rm
  NUV}$ values consists of the NUV-luminous stars
(designated by blue points in Figure~\ref{fig-uv}). The negative values in the
distribution result from the subtraction of a median-fit basal level
when calculating $R'_{\rm NUV}$ (see Equation~\ref{eq-frac}). We
applied a small offset ($R'_{\rm NUV}\sim1.5\times10^{-8}$) to shift
the distribution peak to zero. This was essentially a correction to
our median basal fit (black dashed line in Figure~\ref{fig-uv}), which
was likely skewed to higher $R'_{\rm NUV}$ values because (i) we were
not considering upper limits and (ii) we were only considering stars
with the lowest NUV flux errors, which typically have the highest NUV fluxes.

The shape of the peak in the main panel of
Figure~\ref{fig-volume} is dictated by photometry
errors and the intrinsic width of the basal NUV locus (see
Section~\ref{sec-quiescent}). We needed to extract the extended
tail of NUV-luminous stars for our analysis of the NUVLF of {\it
  young} early-M dwarfs. We did this by reflecting the negative side
of the distribution about the ordinate and subtracting it from the
positive side of the distribution. This essentially removed the
population of stars with basal NUV emission from the
distribution. This assumes that the distribution of $R'_{\rm NUV}$
values in the basal population is symmetric about zero, but does not
assume any distribution in particular. The median \galex{} counts for
basal NUV sources ($\sim$50 counts) were sufficiently high that the
distribution due to \galex{} Poisson errors should be fairly
symmetric. 

This extraction process gave a
residual distribution that represents the NUVLF of young early-M
dwarfs, shown as the blue dashed line in the inset of Figure~\ref{fig-volume}
and tabulated in Table~\ref{tab-lf}. Errors were found using a standard
bootstrap method. We sampled with replacement from the original set of $R'_{\rm
  NUV}$ values until we obtained a bootstrap sample that contained the
same number of data points as the original sample. We then
re-constructed the NUVLF following the
same steps as above, but instead using the bootstrap sample. We
repeated this 100 times and used the standard deviation in each
bin as an estimate of our errors.

\bigskip
\bigskip

\subsection{Including Upper Limits} \label{sec-avni}

Most of the LG11 stars without matches in the \galex{} AIS catalog (2226 out
of 3622) are either late-M dwarfs (and therefore not considered in this
study) or stars within 20 degrees of the Galactic plane (where \galex{}
AIS coverage is sparse; \citealt{2014AdSpR..53..900B}). To determine
which of the remaining 1396 LG11 stars were true non-detections (i.e.,
stars located in an area of sky observed by \galex{} AIS but
too faint to be detected), we searched for \galex{} AIS tiles covering
their coordinates using the aforementioned {\it GalexView} online tool
(see Section~\ref{sec-uvsample}). We only
considered \galex{} AIS tiles with centers less than 0.5 degrees from
the LG11 coordinates of the candidate non-detections. This
separation limit ensured that the stars would have actually been
located on the tile, but not on the tile edge where
photometry can be significantly degraded
\citep{2014AdSpR..53..900B}. This search resulted in the identification of
638 non-detections for which we estimated upper limits.

To calculate upper limits, we derived a relation between
\galex{} exposure time ($t_{\rm NUV}$) and limiting NUV magnitude
($m_{\rm lim}$) using the online \galex{} Exposure Time
Calculator.\footnote{http://sherpa.caltech.edu/gips/tools/expcalc.html}
We queried the tool for exposure times given various $m_{\rm NUV}$ values
and required ${\rm SNR}\sim2$. We used a hypothetical star with ${\rm
  T_{\it eff}}=3500$ K and coordinates $\alpha=240^{\circ}$ and $\delta=-11^{\circ}$
(i.e., an early-M dwarf with median LG11 declination and located away from
the Galactic plane). To check the relation, we plotted $m_{\rm NUV}$
vs. $t_{\rm NUV}$ for the 5267 \galex{}-detected LG11 stars; as
expected, the derived relation between $t_{\rm NUV}$ and $m_{\rm
 lim}$ corresponded to the lower (NUV-dim) bound of the
NUV-detected population. We obtained $m_{\rm lim}$ values for the
638 non-detections by using the $t_{\rm NUV}$ values of their
associated \galex{} AIS tiles with our derived relation. 

We considered these upper limits in our NUVLF using the
method of \cite{1980ApJ...238..800A}. This method employs a
non-parametric, recursive approach to statistically account for upper
limits when constructing a luminosity function \citep[see Equation 6
  in][]{1980ApJ...238..800A}. As before, we replaced the number of
stars in each $R'_{\rm NUV}$ bin with the sum of their $1/V_{\rm max}$ weights in
order to account for survey biases toward brighter stars. However,
because we were considering \galex{} upper limits
with this method, we relaxed any \galex{} constraints on $V_{\rm
  max}$ by only considering the detection distance limits imposed by
LG11 when calculating $V_{\rm max}$. We again had to apply a small offset ($R'_{\rm
  NUV}\sim6.5\times10^{-9}$) to shift the distribution peak to zero;
this correction was much smaller than before, most likely because
we are now taking into account upper limits. The resulting NUVLF is
shown as the red line in the insert of Figure~\ref{fig-volume} and tabulated in
Table~\ref{tab-lf} with errors calculated using the same bootstrap
method described in Section~\ref{sec-lf}.

\capstartfalse
\begin{deluxetable}{lrr}
\centering
\tablewidth{0pt}
\tablecaption{Near-Ultraviolet Luminosity Functions \label{tab-lf}}
\tablecolumns{3}
\tablehead{
   \colhead{$\log(R^{\prime})$}
 & \colhead{$\rho_{V_{max}}$\textsuperscript{a}}
 & \colhead{$\rho_{\rm Avni}$\textsuperscript{b}} \\
 \colhead{}
 & \colhead{($10^{-5}$ stars pc$^{-3}$ dex$^{-1}$)}
 & \colhead{($10^{-5}$ stars pc$^{-3}$ dex$^{-1}$)}
 }
\startdata
-7.50 &     36.1 $\pm$  41.6 &     14.8 $\pm$  28.4 \\
-7.30 &     55.1 $\pm$  52.0 &     94.9 $\pm$  35.9 \\
-7.10 &     77.3 $\pm$  78.7 &     86.6 $\pm$  40.3 \\
-6.90 &     91.6 $\pm$  74.9 &    134.6 $\pm$  38.6 \\
-6.70 &    200.1 $\pm$  65.6 &    144.2 $\pm$  42.6 \\
-6.50 &    155.2 $\pm$  58.8 &    104.9 $\pm$  39.0 \\
-6.30 &    163.9 $\pm$  52.1 &     76.3 $\pm$  33.7 \\
-6.10 &    154.4 $\pm$  37.5 &    126.6 $\pm$  24.3 \\
-5.90 &    208.3 $\pm$  33.3 &    205.1 $\pm$  19.1 \\
-5.70 &    176.1 $\pm$  27.5 &    176.1 $\pm$  16.5 \\
-5.50 &     57.9 $\pm$  21.4 &     57.9 $\pm$  10.3 \\
\enddata
\tablenotetext{a}{NUVLF from Section~\ref{sec-lf} (no upper limits) \\
\textsuperscript{b} NUVLF from Section~\ref{sec-avni} (with upper limits)}
\end{deluxetable}
\capstartfalse

\section{DISCUSSION} \label{sec-diss}

\subsection{Uncertainties and Sensitivities \label{sec-varyfp}}

The two principal uncertainties in our derivation of the NUVLF are: (i)
the overall FP rate, $p$, which we used to statistically
correct for FPs when constructing the NUVLF (Section ~\ref{sec-fpr}); and (ii)
the distances used to compute the limiting detection volumes,
$V_{max}$, which we used to weight each star's contribution to the
NUVLF (Section ~\ref{sec-lf}). We address these two issues below.

Our derived overall FP rate of $p\sim16$\% is consistent
with the $\sim$16\% spectroscopic binary rate among
nearby  X-ray luminous M dwarfs found by
\citet{2009ApJ...699..649S}. One might expect our FP rate to be
higher as our definition of a FP encompasses additional, wider
binaries. However \cite{2009ApJ...699..649S} selected their sample
based on X-ray fluxes from the {\it ROSAT} All Sky Survey, which was
less sensitive to active stars than the \galex{} AIS. Thus their
sample was more biased toward the most active objects, which
likely have higher FP rates (see below). Agreement between our FP
rate and that of \citet{2009ApJ...699..649S} may therefore simply be a coincidence
resulting from several factors.

\capstartfalse
\begin{figure}
\begin{centering}
\includegraphics[width=8.5cm]{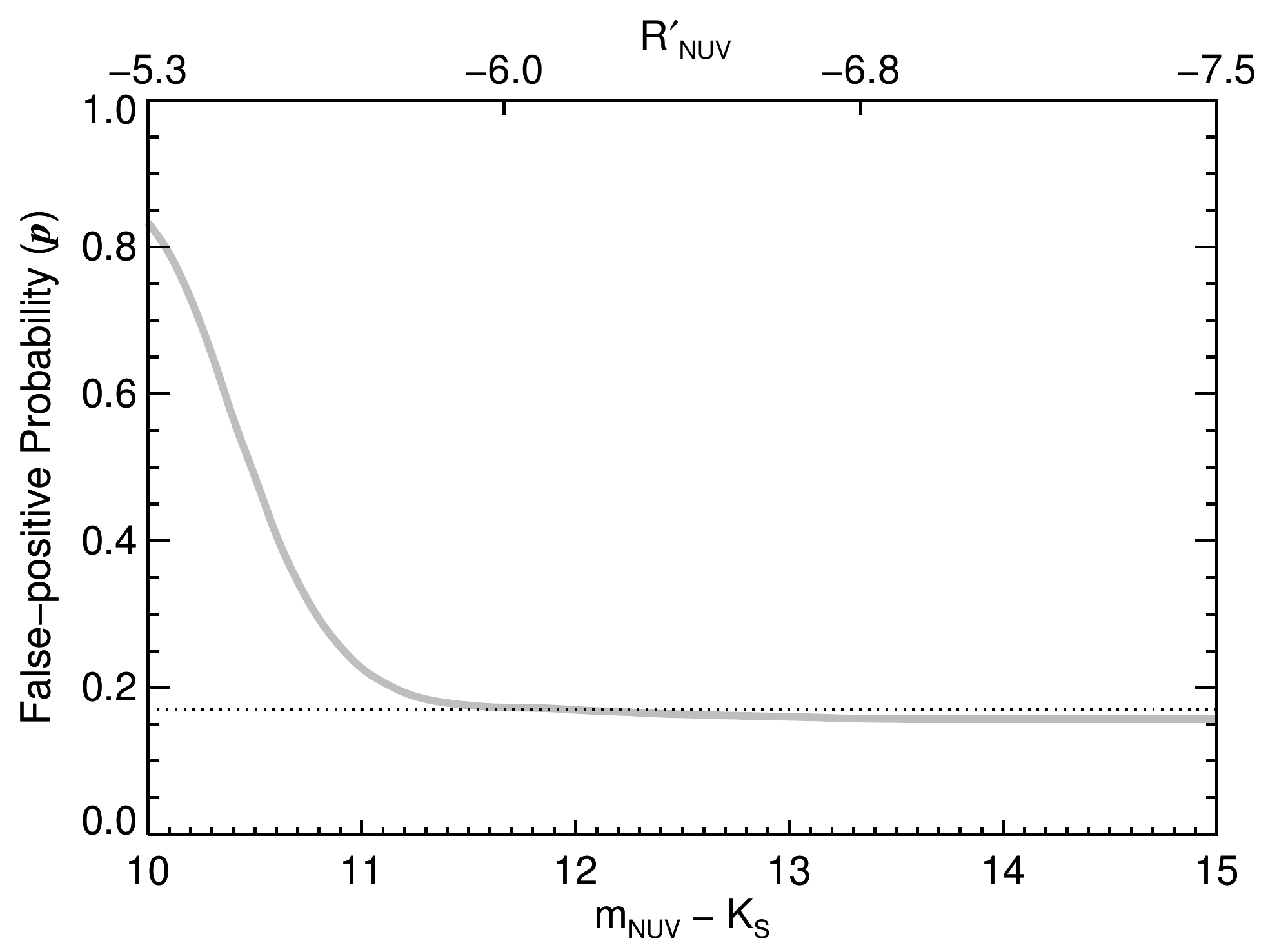}
\caption{\small Varying FP rate among our NUV-luminous sample as a
  function of $m_{\rm NUV}-K_{\rm S}$ and $R'_{\rm NUV}$. The FP rate
  remains constant at $\sim$16\% (i.e., the overall FP rate found in
  Section \ref{sec-fpr}) until $m_{\rm NUV}-K_{\rm S}\sim11$, then
  rises to $>80$\% when considering only the most NUV-luminous M dwarfs.}
\label{fig-mlv}
\end{centering}
\end{figure}
\capstartfalse

The NUV luminosities in our sample span several dex (see
Figure~\ref{fig-volume}). Thus a single, overall FP
rate may be insufficient to describe the entire population, as the most
NUV-luminous sources may have significantly higher FP rates. We tested this by
progressively removing the dimmest NUV sources from our sample, then
repeating the maximum likelihood estimation in Section
\ref{sec-fpr} to re-derive $p$ based on the cropped sample. Results
are given in Figure~\ref{fig-mlv}, which shows a constant
FP rate of $p\sim16$\% (i.e., our overall FP
rate) until $m_{\rm NUV}-K_{\rm S}\sim11$. At higher NUV
luminosities, the FP rate steadily increases, reaching $p\sim80$\%
by $m_{\rm NUV}-K_{\rm S}\sim10$. To test the implications of this
result, we re-derived our NUVLF using this varying FP rate (instead of the constant
$p\sim16$\% value) when multiplying the $1/V_{\rm max}$ weights by
$1-p$ to statistically account for FPs (see Section~\ref{sec-lf}). We
did not expect significant changes to the NUVLF, as there are few
stars in our sample with high enough NUV luminosities to require FP
rates that are significantly larger than the overall FP rate. The
results are compared to the original NUVLF in the inset of
Figure~\ref{fig-volume}. The consistency between the two NUVLFs
suggests that our derivation is not particularly sensitive to this
varying FP rate. Still, our use of a single FP rate is a gross simplification of reality where
there are multiple, sometimes unrelated sources of FPs, each of which
cannot be detected by all our methods applied in
Section~\ref{sec-newfp}.  A more rigorous approach would use a modified version of
Equation \ref{eq-ml2} to estimate multiple $p$ values, one for each FP
source, and also account for the completenesses of each method for each
FP source. However the outcome of the sensitivity analysis
described above suggests that our results would not change significantly.

Another source of uncertainty in our derivation of the NUVLF
is our estimation of stellar distances, which we used to calculate
$V_{max}$ and thus the weighted contribution of each star
to the NUVLF. For most sources we used a $J$-band photometric distance,
where M$_{J}$ was estimated from $V-J$ color using Equation 22
in \cite{2013AJ....145..102L}. However we substituted more accurate
parallax distances for the 561 stars in our sample that also had
trigonometric parallax measurement with errors $<10$\%. We used the
average difference between these photometric and parallax
distances to estimate fractional errors as a function of stellar
distance. These ranged from $\sim$50\% for distances of
$\sim$3 pc (the minimum distance in our sample) to $\sim$10\% for
distances of $\sim$40 pc (the distance containing 95\% of our
sample). We then tested how these distance uncertainties impacted our derived
NUVLF using a Monte Carlo approach. We perturbed each distance by a
random Gaussian deviate scaled to our estimated fractional errors, then re-ran our
derivation of the NUVLF. We repeated this 100 times and took the
standard deviation in each luminosity bin as an estimate of the impact
of our distance uncertainties on our NUVLF. The impact appeared to be
negligible: the variation was only $\sim$14\% of
the original NUVLF in each bin, well within the errors estimated by
bootstrap sampling (see Section~\ref{sec-lf}).

\subsection{NUV Age-Activity Relation \label{sec-age}}

Relations between stellar age and emission at high-energy wavelengths are
typically expressed as two-parameter power laws of the form F$_{\lambda}=\alpha
t^{\beta}$ \citep[c.f.,][]{2005ApJ...622..680R} where $t$ is the age
of the star and the two parameters, $\alpha$ and $\beta$, are the zero
point and slope of the power law,
respectively. \cite{2013MNRAS.431.2063S} derived an age-activity relation
for early-M dwarfs by combing their sample of 159 nearby field M dwarfs
(assuming an age of $\sim$3 Gyr) with an additional sample of young
($\sim$1 Myr) M dwarfs from the TW Hya
association. They found that $\beta=-0.84\pm0.08$ at \galex{} NUV
wavelengths. 

Because stellar ages for our sample are mostly unknown, we first attempted
to infer an age-activity relation by fitting our observed
NUVLF to model NUVLFs constructed from assumed power-law
age-activity relations. To create model populations, we assumed a
constant star-formation rate (and thus a uniform age distribution) with a
maximum age of $\sim$10 Gyr \cite[i.e., the approximate age of the Galactic
disk at the present solar radius;][]{2014A&A...565A..89B}. We then used a
two-parameter power law, described above, to assign $R'_{\rm NUV}$
values to each synthetic star based on its model age. We created the
model NUVLF by binning the synthetic population according to the same
$R'_{\rm NUV}$ bins as in the inset of Figure~\ref{fig-volume}, then normalizing
the model distribution such that the integral under the model function
equaled that of the real function. We searched for best-fit parameters by minimizing
reduced $\chi^2$. We calculated reduced $\chi^2$ by taking
the difference between the model and observed NUVLFs at each $R'_{\rm
  NUV}$ bin, applying the errors shown in the inset of Figure~\ref{fig-volume} to
the observations, and dividing by the number of bins minus the
number of power-law parameters.
We found best-fit parameters of $\beta\approx-1.29$ and
$\alpha\approx2.4\times10^{-6}$ with reduced $\chi^{2}\approx6.3$. The
best-fit model is compared to the observed NUVLF in the inset of
Figure~\ref{fig-volume}. Clearly this simplified model is unable to
account for the observed NUVLF, likely due to our two key
assumptions: (i) our neglect of a constant (i.e., saturated) level of 
NUV emission at very young ages and/or (ii) our assumption of a constant star-formation rate.
We discuss (i) below and (ii) in Section~\ref{sec-sfr}.

\capstartfalse
\begin{figure}
\begin{centering}
\includegraphics[width=8.5cm]{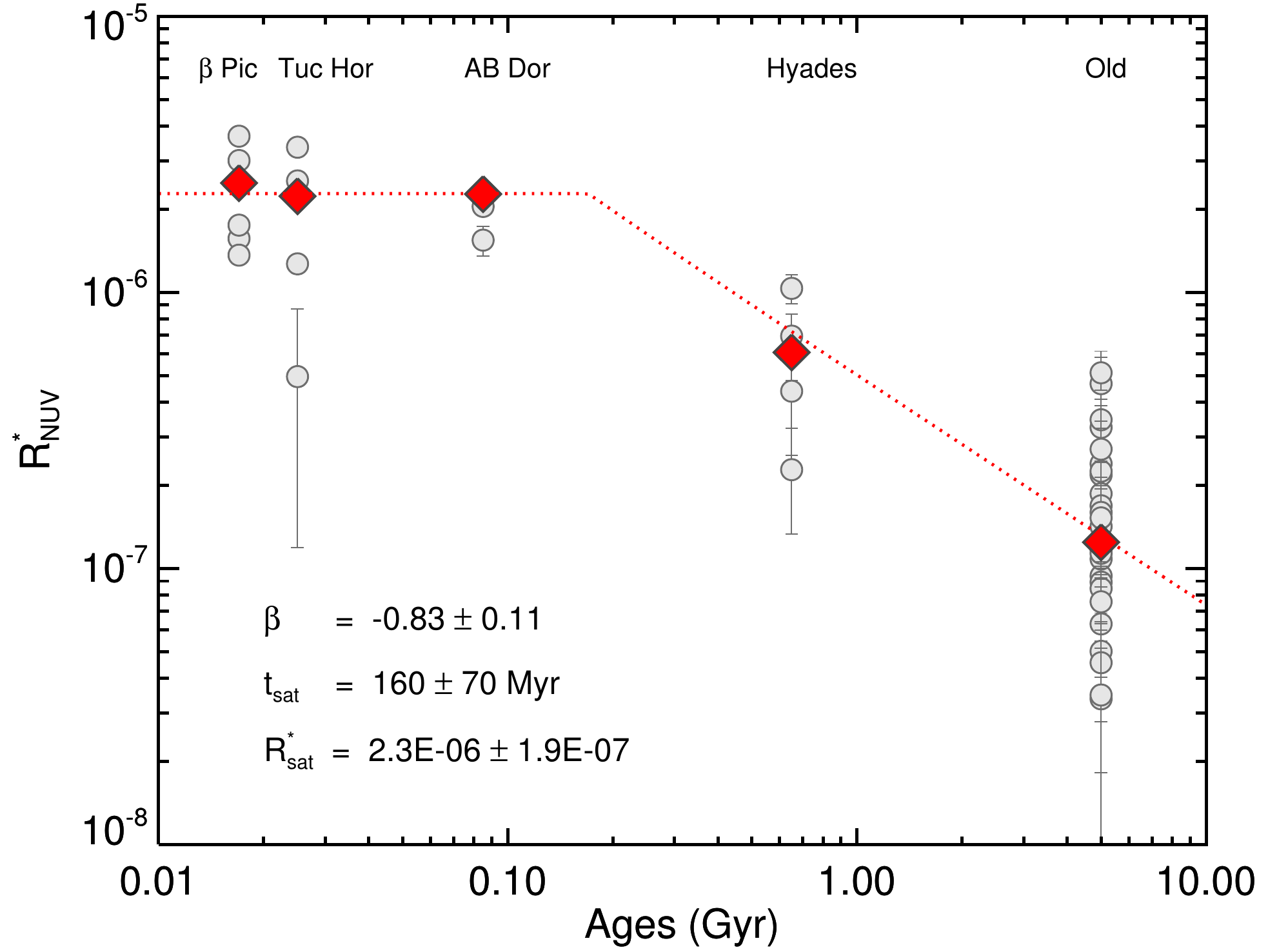}
\caption{\small Photosphere-subtracted fractional NUV luminosity, $R^{*}_{\rm NUV}$, as
  a function of age for the 20 YMG members in our sample as well as 28
  old field stars identified by their high space motions. Red diamonds are
  median $R^{*}_{\rm NUV}$  values for each age group. The best-fit
  broken power law, shown by the dotted red line with parameters
  printed for reference, was found using $\chi^2$ minimization and
  bootstrap sampling (see text for details).}
\label{fig-ymg}
\end{centering}
\end{figure}
\capstartfalse

Saturated (i.e., constant) NUV emission in young early-M dwarfs was
recently reported by \cite{2014AJ....148...64S}. They used early-M members of nearby
young moving groups (YMGs) to derive a NUV age-activity
relation that showed NUV emission remaining constant for
$\approx300$ Myr then declining as a power law
with $\beta=-0.84\pm0.09$. This power law index agrees with the
previous findings of \citet{2013MNRAS.431.2063S}, who did not consider
saturated NUV emission due to the lack of stars younger than a few hundred
Myr in their sample. We therefore used the YMG members in
our sample to empirically derive a power-law age-activity relation
that included a saturation component. We first searched the literature
to identify 32 candidate YMG members in our sample. We then required
high ($\ge95$\%) membership probabilities, which we obtained using the
BANYAN \citep[Bayesian Analysis for Nearby Young
AssociatioNs;][]{2013ApJ...762...88M} online tool.\footnote{http://www.astro.umontreal.ca/$\sim$malo/banyan.php} For
the Hyades, which is not included in BANYAN, we used the approach of
\cite{2014AJ....148...64S} by requiring a kinematic link to the
YMG. We also removed any FPs identified in Section~\ref{sec-fp}. This
resulted in a final set of 20 YMG members, which are flagged in Table~\ref{tab-params}.

In order to extend our age-activity relation to older field stars, as
well as make our derived age-activity relation comparable to
previous works, we re-defined our fractional NUV
luminosity ($R'_{\rm NUV}$; see Equation~\ref{eq-frac}). Instead of
subtracting the observed basal NUV luminosity (L$_{\rm basal}$) we subtracted the
photospheric NUV luminosity predicted by PHOENIX models (L$_{\rm
  phot}$) to obtain $R^{*}_{\rm NUV}$ (the predicted values of
L$_{\rm phot}$ as a function of $V-J$ are shown by the dash-dotted
line in Figure~\ref{fig-uv}). The $R^{*}_{\rm NUV}$ values for our
20 YMG members are plotted as a function of their age in
Figure~\ref{fig-ymg}. Also shown are 28 older field
stars from our sample, which were identified by having space
motions $>1\sigma$ from the mean of active/young stars in at least two
spatial dimensions \cite[space motion information was obtained
from][]{2013AJ....145..102L}. Similar to the findings of
\cite{2014AJ....148...64S}, we observed a saturated level of NUV
emission lasting a couple Myr, followed by a power-law decline. 

To derive our age-activity relation, we fit a broken power law to the
median $R^{*}_{\rm NUV}$ values of each age
group. To obtain the best-fit parameters and associated errors,
we constructed 100 bootstrap samples from our data, found the model
parameters associated with the minimum reduced $\chi^{2}$ for each
bootstrap sample, then took the mean and standard deviation. The
results are shown in Figure~\ref{fig-ymg},
indicating a saturated NUV emission level at $R^{*}_{\rm
  NUV}=2.3\pm0.19\times10^{-6}$ until $160\pm70$ Myr of
age, after which NUV emission declines as a power law with
slope $\beta=-0.83\pm0.11$. Our $\beta$ value agrees well with
those found by \cite{2013MNRAS.431.2063S} and
\cite{2014AJ....148...64S}. Our saturation timescale is also
consistent with that of \cite{2014AJ....148...64S}, although their
value of 300 Myr is slightly higher, possibly due to our
lack of data points between 100 and 600 Myr. 

\subsection{Inferred Age Distribution of Early M Dwarfs \label{sec-sfr}}

\begin{figure}
\begin{centering}
\includegraphics[width=8.5cm]{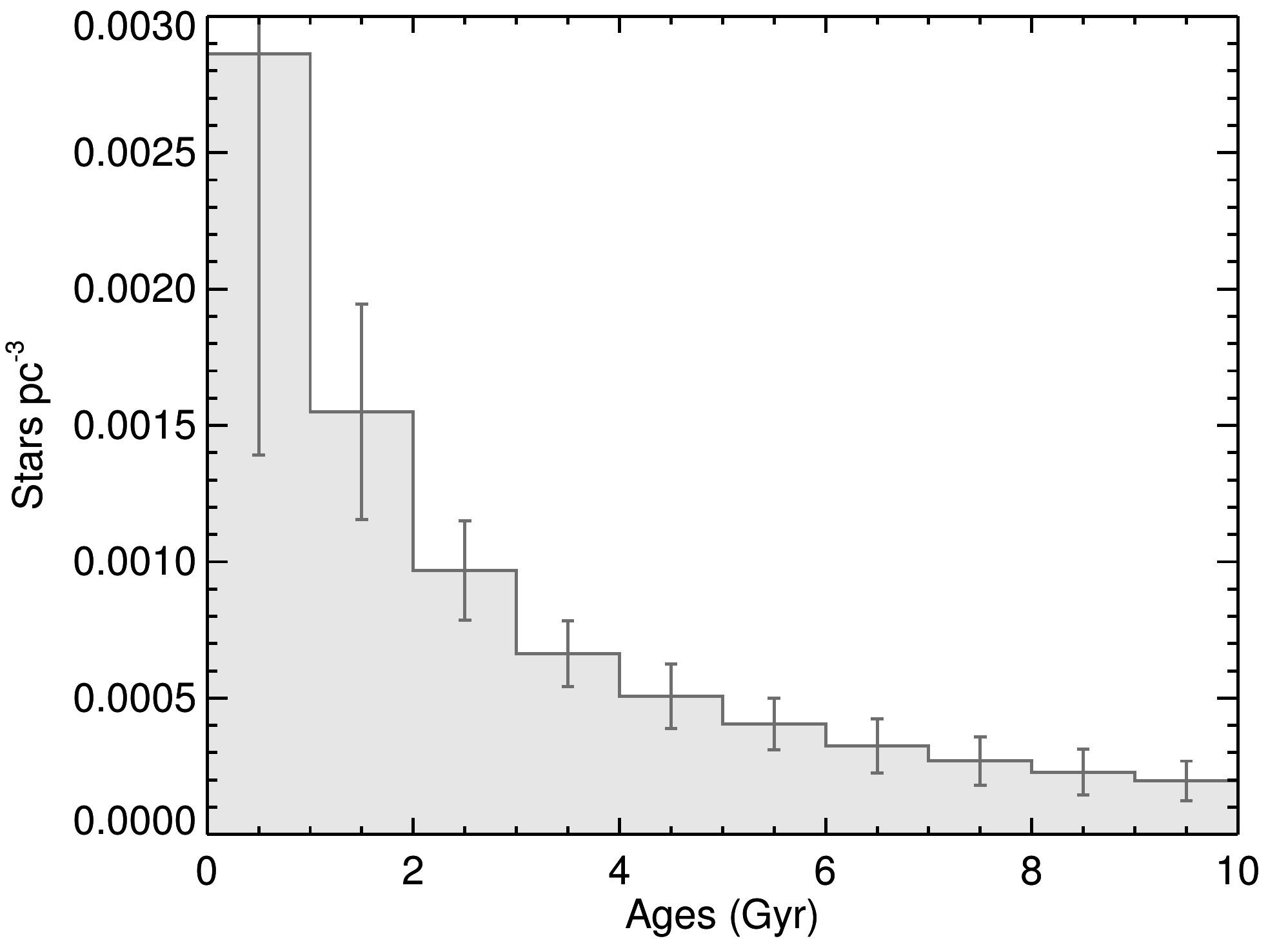}
\caption{\small The age distribution for our sample, constructed by
  applying our derived age-activity relation found in
  Section~\ref{sec-age}. The large error bars at young ages reflect the sensitivity
  of the age distribution to the input parameters when they are varied
  according to their uncertainties.}
\label{fig-sfr}
\end{centering}
\end{figure}

We investigated possible star-formation rate histories by
deriving an age distribution for our sample using a Monte Carlo
approach. For each star we perturbed its $R^{*}_{\rm NUV}$ value by a random
Gaussian deviate scaled to its error on $R^{*}_{\rm NUV}$. We
also perturbed each parameter of the age-activity relation derived in
Section~\ref{sec-age} in an analogous manner. We used these perturbed values
to estimate an age for each star and then constructed an age
distribution by summing the $1/V_{\rm max}$ weights of the stars in each
age bin. We repeated this 100 times then summed the normalized
distributions to create the final age distribution shown in
Figure~\ref{fig-sfr}. We found that at young ages our derived age
distribution varied greatly depending on the input parameters,
resulting in large errors and thus an uncertain distribution at young ages.

There has been much discussion on the star-formation rate history of the
Solar neighborhood. \cite{2002AJ....123.3356G} used a spectroscopic
survey of 676 nearby M dwarfs to infer a constant star-formation rate
over the last 4 Gyrs. However, there have been several studies (which
do not utilize M dwarfs) that indicate elevated star-formation rates in recent history.
\cite{2000MNRAS.316..605H} used {\it Hipparcos} data to
claim rapidly fluctuation star-formation rates with frequencies of $\sim$0.5
Gyrs, however our data does not have sufficient time
resolution to be compared to their work. \cite{2011MNRAS.415.2827B}
used a sample of 442 star clusters within 1 kpc to show a recent
(220-600 Myrs) local burst in star formation that is twice the average star-formation
rate. \cite{2014ApJ...791...92T} used the luminosity function of
white dwarfs within 20 pc to show enhanced star-formation rates
within the last 5 Gyrs compared to that of $5-10$ Gyrs. The latter
two studies appear to be most consistent with our Figure~\ref{fig-sfr}, if the
slope in our age distribution can be considered significant.

\capstartfalse
\capstartfalse

\subsection{Implications for Habitability \label{sec-hab}}

Our analysis indicates an era of saturated NUV emission for young M
dwarfs lasting $\sim$100--200 Myr (although possibly up to $\sim$$300$
Myr; see \citealt{2014AJ....148...64S}). Roughly 120 stars in our
sample have sufficiently high NUV luminosities ($R^{*}_{\rm
  NUV}\gtrsim2.3\times10^{-6}$ or $m_{\rm NUV}-K_{\rm S}\lesssim10.9$)
to place them in this saturation age interval. Correcting for a FP
rate of $\sim$$25\%$ (see Figure~\ref{fig-mlv}) reduces this saturated
sample to $\sim$90 stars or $\sim$$2\%$ of our sample. Any planets
in the habitable zones of these young M dwarfs will be exposed
to persistent, elevated NUV irradiation. Because the dissociation energies of
several key atmospheric molecules are in the NUV (e.g., H$_{2}$O at
2398\AA, CO$_{2}$ at 2247\AA, CH$_{4}$ at 2722\AA), the atmospheres of these planets
can be significantly altered by photodissociation. Although detailed studies of
these processes are just beginning, recent results suggest
the implications may be significant (e.g., see
\citealt{2014arXiv1410.2112M} for the effects of Lyman $\alpha$ radiation
on the atmosphere of mini-Neptune GJ 436b, which orbits an M3 star).

The saturation timescale of a few hundred Myr also roughly corresponds to the
final ``giant impact" phase predicted by
terrestrial planet formation models. This interval scales with
orbital period, which is proportionally shorter for planets in the
compact habitable zones of M dwarfs \citep{2012AREPS..40..251M}. This
coincidence implies that the early atmospheres of planets around young M
dwarfs are subject to erosion via heat injection from impactors in
addition to NUV irradation by their host stars.  

We also find that the vast majority of stars in our saturated sample
($\sim$70\% after FP correction) have F$_{\rm FUV}$/F$_{\rm
  NUV}\ge0.1$, which is at least two orders of magnitude above the solar value of
$\sim$0.001 (see also \citealt{2013ApJ...763..149F}). This raises the
potential for high rates of {\it abiotic} atmospheric O$_{2}$ and
O$_{3}$ (produced from CO$_{2}$)---two molecules that have been
proposed as biosignatures on Earth-like planets \cite[see][and references
therein]{2014E&PSL.385...22T}. Moreover, even some of the older M dwarfs in
our sample, which exhibit basal levels of NUV emission above model-predicted
photospheric values (see Figure~\ref{fig-uv}), have F$_{\rm FUV}$/F$_{\rm
  NUV}\ge0.1$. Thus very blue light may remain an important
consideration for habitable-zone planets around these very red stars. 

\bigskip

\section{SUMMARY}\label{sec-sum}

We have constructed a NUV luminosity function for young,
early-M dwarf stars. We cross-correlated the \cite{2011AJ....142..138L} 
catalog of bright M dwarfs with the \galex{} all-sky
catalog of NUV sources to identify a sample of 4805 NUV-detected
early-M dwarfs (M0-M3). Of these, 797 had NUV emission significantly ($>2.5\sigma$) in
excess of an observed basal emission level; parameters of these
candidate young stars are summarized in Table~\ref{tab-params}. When
constructing the NUV luminosity function from this sample, we
corrected for false positives (i.e., systems appearing NUV-luminous
for reasons other than stellar youth; Section~\ref{sec-fp}), the
biases of the source catalogs toward intrinsically brighter sources
(Section~\ref{sec-lf}), and \galex{} upper limits
(Section~\ref{sec-avni}). Key findings from our analysis are:

\begin{itemize}

  \item Plotting $V-J$ (a proxy for stellar effective temperature)
    vs. $m_{\rm NUV}-K_{\rm S} $ (a measure of NUV luminosity) for our
    sample of 4805 NUV-detected early-M dwarfs shows two distinct
    populations. The majority of sources fall along a locus, but about
    20\% of the sample appears NUV luminous
    with $m_{\rm NUV}-K_{\rm S} $ colors at least 2.5$\sigma$
    ($\sim$1.12 magnitudes) bluer than the main locus.

  \item All sources in our sample appear to exhibit a basal level
    of NUV emission above the expected photospheric value predicted by
    atmospheric models. This basal level of NUV emission for all M
    dwarfs regardless of age was first noted by
    \cite{2013MNRAS.431.2063S}. Our empirical fit to this basal level
    of NUV emission as a function of $V-J$ color is $m_{\rm NUV}-K_{\rm S} = 7.72 + 1.66 (V-J)$.

  \item We conducted an extensive search for false positives (i.e., systems appearing
    NUV-luminous for reasons other than stellar youth) using
    medium-resolution optical spectra, high-resolution adaptive optics imaging,
    time-series photometry, and literature searches. We applied a
    maximum likelihood scheme to estimate the overall occurrence of
    false positives in our NUV-luminous sample to be $\sim$16\%. However,
    we also found that this false-positive rate is significantly
    higher for the most NUV-luminous sources, reaching $\sim$80\% by
    $m_{\rm NUV}-K_{\rm S}\sim10$. 

  \item We derived a NUV luminosity function for young, early-M dwarfs
    that was corrected for false positives, the biases of the source catalogs toward
    intrinsically brighter sources, and \galex{} upper limits.
    Our derived NUV luminosity function is {\it inconsistent} with
    predictions from a constant star-formation rate and age-activity
    relation described by a two-parameter power law.

  \item We derived a NUV age-activity relation using the 20
    YMG members in our sample with known ages as well as 28 older field
    stars identified by their high space motions. Results indicate a
    saturated NUV emission level for young, early-M dwarfs until
    $160\pm70$ Myr of age, after which NUV emission declines
    with a power-law slope of $-0.83\pm0.11$ (consistent with
    \citealt{2014AJ....148...64S} and \citealt{2013MNRAS.431.2063S}). 
    However, because even the oldest stars in our sample exhibit
    basal levels of NUV emission above predicted photospheric
    values, this power-law decline in NUV emission is likely only
    applicable to a few Gyr of age.

\end{itemize}

\section{ACKNOWLEDGEMENTS} \label{sec-ack}

MA and EG acknowledge support from NASA grants NNX10AQ36G
(Astrobiology: Exobiology \& Evolutionary Biology) and NNX11AC33G
(Origins of Solar Sytem). This research utilized the NASA Astrophysics
Data System, SIMBAD database, and Vizier catalogue
access tool. It also made use of the AAVSO Photometric All-Sky Survey,
funded by the Robert Martin Ayers Sciences Fund. We thank the
dedicated staff of the UH88, MDM, SAAO, and CASLEO observatories from
which many spectra were obtained for this work. We especially thank
Greg Aldering for years of assistance with SNIFS on the UH88. We also
thank Evgenya Shkolnik for her very useful comments regarding \galex{}. This paper
used observations taken with the Robo-AO system. The Robo-AO system is
supported by collaborating partner institutions, the California
Institute of Technology and the Inter-University Centre for Astronomy
and Astrophysics, by the National Science Foundation under Grant
Nos. AST-0906060, AST-0960343, and AST-1207891, by a grant from the
Mt. Cuba Astronomical Foundation and by a gift from Samuel
Oschin. C. B. acknowledges support from the Alfred P. Sloan Foundation.

\newpage

\bibliography{bib_eg.bib}

\newpage

\capstartfalse
\begin{sidewaystable}
\centering
\tiny
\caption{Parameters of UV-luminous Early-M Dwarfs\label{tab-params}}
\begin{tabular}{lcccrrcrccccccc}
Star & $V$ & $J$ & $m_{\rm NUV}-K_{\rm S}$ & $F_{\rm NUV}$ ($\mu$Jy)  & $F_{\rm FUV}$ ($\mu$Jy) & $F_{\rm X}$ (ergs cm$^{-2}$ s$^{-1}$) & EW$_{\rm H \alpha}$ (\AA) & $\log(R^{\prime}_{\rm NUV})$ & Robo-AO\textsuperscript{a} & Low H$\alpha$\textsuperscript{b} & Shifted H$\alpha$\textsuperscript{c} & SuperWASP\textsuperscript{d} & FP?\textsuperscript{e} \\
\hline
 PM\_I00001+6943 & 13.52 & 9.70 & 10.95 &    44.04 &   99.99 & 2.95E-13 & 99.99 & -5.67 & --- & --- & --- & --- &   N   \\
 PM\_I00024-4601 & 12.43 & 9.18 & 11.98 &    27.05 &    5.78 &    99.99 & 99.99 & -6.22 & --- & --- & --- &  No &   N   \\
 PM\_I00059+4129 & 12.96 & 9.40 & 11.56 &    33.08 &   10.22 &    99.99 &  1.14 & -7.48 &  No &  No &  No & Yes &   D   \\
 PM\_I00072-1036 & 11.85 & 8.83 & 11.29 &    70.18 &   11.78 & 4.48E-13 &  0.80 & -5.89 & --- &  No &  No & --- &   N   \\
 PM\_I00107-2039 & 13.39 & 9.48 & 11.18 &    43.77 &   10.82 & 9.27E-13 & 99.99 & -5.77 & --- & --- & --- & --- &   N   \\
 PM\_I00117-1139 & 12.67 & 9.85 & 11.05 &    34.89 &   99.99 &    99.99 & -0.04 & -5.83 & --- &  No & --- & --- &   N   \\
 PM\_I00118-5521 & 13.34 & 9.32 & 11.50 &    36.21 &   18.78 &    99.99 &  3.18 & -5.86 & --- &  No & --- & --- &   N   \\
PM\_I00162+1951W & 11.79 & 7.88 & 11.64 &   117.08 &   22.29 & 2.39E-12 &  6.16 & -5.90 & --- &  No & --- & --- &   N   \\
 PM\_I00166+3000 & 12.79 & 9.11 & 12.42 &    20.21 &    3.56 & 1.02E-13 & 99.99 & -6.38 & --- & --- & --- & --- &   N   \\
 PM\_I00197-2233 & 13.77 & 9.88 & 12.44 &     9.59 &   99.99 &    99.99 & 99.99 & -6.33 & --- & --- & --- & --- &   N   \\
 PM\_I00206-5340 & 13.80 & 9.68 & 13.30 &     5.18 &   99.99 &    99.99 & 99.99 & -6.71 & --- & --- & --- & --- &   N   \\
 PM\_I00211+4456 & 13.62 & 9.60 & 10.89 &    52.93 &   99.99 & 1.07E-12 &  5.20 & -5.67 &  No &  No &  No & --- &   N   \\
 PM\_I00216-4605 & 12.21 & 8.32 & 11.14 &   133.11 &   28.10 & 7.46E-12 &  1.94 & -5.76 & --- &  No & --- &  No &   N   \\
 PM\_I00234+2418 & 13.03 & 9.75 & 11.01 &    40.74 &   99.99 &    99.99 & -0.31 & -7.48 & Yes &  No & --- &  No &   D   \\
PM\_I00235+0947S & 12.75 & 9.79 & 10.59 &    56.81 &   10.92 &    99.99 & -0.77 & -7.48 &  No &  No & Yes & --- &   D   \\
 PM\_I00235+2014 & 11.00 & 8.14 & 10.84 &   193.75 &   99.99 & 1.35E-12 &  0.30 & -7.48 &  No &  No & Yes &  No &   D   \\
 PM\_I00241-6211 & 11.33 & 8.38 & 11.00 &   144.75 &   32.66 & 1.04E-12 &  0.94 & -7.48 & --- &  No & --- & --- &   L   \\
 PM\_I00245-2522 & 13.62 & 9.84 & 11.29 &    29.08 &    5.05 & 3.07E-13 &  2.68 & -5.84 & --- &  No &  No &  No &   N   \\
 PM\_I00250-3646 & 12.46 & 8.64 & 12.00 &    42.38 &   12.20 & 9.64E-13 &  1.12 & -7.48 & --- &  No & --- &  No &   L   \\
\hline
\end{tabular}
\medskip \\
\raggedright
\textsuperscript{a}Yes = found as FP by Robo-AO FP detection technique (Section 4.2.1); No  = not found as FP by Robo-AO FP detection technique;  --  = not observed by Robo-AO. \\
\textsuperscript{b}Same as (a) but for missing H$\alpha$ FP detection technique (Section 4.2.2). \\
\textsuperscript{c}Same as (a) but for shifted H$\alpha$ FP detection technique (Section 4.2.3). \\
\textsuperscript{d}Same as (a) but for SuperWASP FP detection technique (Section 4.3.4). \\
\textsuperscript{e}L = found as FP in literature (Section 4.1); D = found as FP using detection techniques (Section 4.2); N = not found as FP in literature or by FP detection techniques; R = removed from NUVLF (Section 5.2); Y = YMG member used in age-activity relation derivation (Section 6.2). \\
Table 4 is published in its entirety in the electronic edition of \emph{The Astrophysical Journal}. A portion is shown here for guidance regarding its form and content.
\end{sidewaystable}
\capstartfalse

\end{document}